\documentclass[nofootinbib,prd,aps,superscriptaddress,preprintnumbers]{revtex4}
\pdfoutput=1
\usepackage{amsmath,amssymb,euscript}
\usepackage{color}
\usepackage{accents}
\usepackage{hyperref}
\usepackage{ulem}
\usepackage{epsfig}
\usepackage{varioref}
\usepackage{xcolor}
\usepackage{verbatim}

\def\beq{\begin{equation}}
\def\eeq{\end{equation}}

\renewcommand{\emph}{\textit}

\graphicspath{{figs/}}

\begin{document}

\begin{flushright}

\end{flushright}

\title{Virtual signatures of dark sectors in Higgs couplings}

\author{Alexander Voigt}
\affiliation{Institute for Theoretical Particle Physics and Cosmology, RWTH Aachen University, D-52074 Aachen, Germany}
\author{Susanne Westhoff}
\affiliation{Institute for Theoretical Physics, Heidelberg University, D-69120 Heidelberg, Germany}

\vspace{1.0cm}
\begin{abstract}
\vspace{0.2cm}\noindent 
Where collider searches for resonant invisible particles loose steam, dark sectors might leave their trace as virtual effects in precision observables. Here we explore this option in the framework of Higgs portal models, where a sector of dark fermions interacts with the standard model through a strong renormalizable coupling to the Higgs boson. We show that precise measurements of Higgs-gauge and triple Higgs interactions can probe dark fermions up to the TeV scale through virtual corrections. Observation prospects at the LHC and future lepton colliders are discussed for the so-called singlet-doublet model of Majorana fermions, a generalization of the bino-higgsino scenario in supersymmetry. We advocate a two-fold search strategy for dark sectors through direct \emph{and} indirect observables.
\end{abstract}
\maketitle


\section{Introduction}
\noindent Dark matter at the weak scale is under pressure. ``Dark'' particles, i.e., new particles with weak interactions with the standard model (SM), have been searched for extensively at colliders, as well as in direct and indirect detection experiments. The absence of any clear evidence suggests that either the hypothesis of weak-scale dark matter needs to be rethought or it is simply eluding observation. Yet, dark particles with masses in reach of high-energy colliders are well motivated candidates of thermal relics~\cite{Srednicki:1988ce,Kolb:1990vq,Gondolo:1990dk}. Beyond dark matter, sectors of dark particles with Higgs boson interactions can facilitate electroweak baryogenesis~\cite{Anderson:1991zb,Grojean:2004xa}. We therefore take up the position that a dark sector around the weak scale might very well exist. But we need to extend current searches to cover the full range of its possible manifestations.

Focusing on high-energy colliders, LHC searches for missing energy from resonant dark particles have already tested some of the theory space of dark sectors~\cite{atlas-cms:dm-searches}. However, missing energy searches can loose their power if dark particles are heavy and/or their production cross section is small, if visible decay products are soft and hard to observe, if SM backgrounds are large and difficult to overcome. In such situations, it is important to explore dark sectors through {\it indirect searches} for virtual dark particles in SM precision observables. Two areas that are and will continue to be under particular scrutiny at the LHC and future lepton colliders (FLC) are the Higgs boson and electroweak interactions. Precise predictions and measurements in these areas allow us turn SM tests into indicators of dark sectors.

The goal of our work is to show that indirect searches for dark sectors can be complementary and sometimes superior to direct searches with missing energy. We will substantiate this statement using the example of a simple model of new Majorana fermions interacting with the SM through the Higgs boson, known as the singlet-doublet fermion Higgs portal. Our model is similar to the bino-higgsino scenario in the minimal supersymmetric standard model (MSSM)~\cite{Jungman:1995df}, except that the couplings are not set by supersymmetry, but free parameters. The singlet-doublet model has received a lot of attention, due to its interesting phenomenology in the context of dark matter~\cite{Mahbubani:2005pt,DEramo:2007anh,Enberg:2007rp,Kearney:2016rng}, baryogenesis~\cite{Carena:2004ha,Chao:2015uoa}, and observation prospects at colliders~\cite{Enberg:2007rp,Calibbi:2015nha,Freitas:2015hsa}. Besides those virtues, it provides naturally strong renormalizable interactions between the dark fermions and the Higgs boson. This feature suggests sizable effects in Higgs observables.

The relevant properties of our dark fermion model are introduced in Section~\ref{sec:model}. In Section~\ref{sec:higgs-df}, we discuss the effects of virtual dark fermions in Higgs couplings, specifically in Higgs interactions with weak gauge bosons and Higgs self interactions. Once the main features of the anomalous Higgs interactions are laid out, we discuss their phenomenology in Higgs observables at the LHC and future lepton colliders in Section~\ref{sec:formfactors}. Complementary probes of dark fermions in electroweak precision measurements will be the topic of Section~\ref{sec:ewpo}. In Section~\ref{sec:vacuum}, we comment on the impact of dark fermions on the stability of the electroweak vacuum, which tends to limit effects in Higgs interactions. In Section~\ref{sec:future}, we explore the reach of indirect searches for dark sectors at the LHC and future high-energy colliders. We compare the results with the sensitivity of resonant searches in Section~\ref{sec:resonant} and conclude in Section~\ref{sec:conclusions}.

\section{A Higgs-portal model of dark fermions}\label{sec:model}
\noindent We investigate a minimal realization of a dark sector with fermions that have Yukawa interactions with the Higgs boson. Our model is a generalization of the bino-higgsino system in the MSSM, without any requirements on the couplings imposed by supersymmetry. The SM is supplemented by a Majorana fermion singlet, $\chi_S$, and two weak fermion doublets, $\chi_D$ and $\chi_D^c$, with hypercharges $+1/2$ and $-1/2$,
\begin{align}\label{eq:fields}
\chi_S=\chi_S^0,\quad \chi_D=(\chi_D^+,\chi_D^0),\quad \chi_D^c = (-\chi_D^{0\ast},\chi_D^-).
\end{align}
Throughout our analysis, $\chi_S$, $\chi_D$, and $\chi_D^c$ denote left-handed Weyl spinors. We assume that the dark fermions have vector-like electroweak interactions, so that the model is free from gauge anomalies. It is furthermore assumed that dark fermions are odd under a discrete $\mathbb{Z}_2$ symmetry, $\chi \to -\chi$, while SM fermions are even. The lightest neutral state is thus a stable DM candidate, and mixing between dark fermions and SM fermions is absent. The mass spectrum in our model is determined by the Lagrangian
\begin{align}\label{eq:maj-sd}
{\cal L}\supset m_D\chi_{D}^c\epsilon\chi_{D} -\tfrac{1}{2}m_S\chi_S\chi_S - y (H^\dagger \chi_{D} \chi_S  - \chi_S  \chi_{D}^c \epsilon H) + \text{h.c.},
\end{align}
where $H=\big(h^+,(v + h + i\eta)/\sqrt{2}\big)$ is the SM Higgs field with a vacuum expectation value $v=246\,\text{GeV}$. The antisymmetric tensor $\epsilon$ acts on the weak $SU(2)$ doublets. We assume that $\chi_D$ and $\chi_D^c$ couple with equal strength and opposite sign to the Higgs field. Two of the three parameters $\{m_D,m_S,y\}$ can always be made real and positive by a redefinition of the fermion fields. We consider the case where also the third parameter is real and choose $m_D,y >0$. After electroweak symmetry breaking, the dark Yukawa coupling $y$ induces a mixing between singlet and doublet fermions, parametrized by an angle $\theta$ with
\begin{align}\label{eq:mixing}
\sin^2\theta = \frac{1}{2} \biggl(1 + \frac{m_D-m_S}{\Delta m}\biggr),\qquad \Delta m = \sqrt{(m_D - m_S)^2 + 4 (yv)^2}.
\end{align}
This mixing yields three physical neutral Majorana fermions $\chi_1^0,\,\chi_2^0,\,\chi_3^0$ and two charged fermions $\chi^\pm$ with masses $m_1$, $m_2$, $m_3$, and $m_c$ given by\footnote{With our parameter choice (above Eq.~\ref{eq:mixing}), $m_1$ and in principle also $m_3$ can be negative for large negative $m_S$ or large $\Delta m$.}
\begin{align}\label{eq:masses}
m_1 & = \tfrac{1}{2}\big(m_D+m_S - \Delta m\big),\quad m_2 = m_D = m_c,\quad m_3 = \tfrac{1}{2}\big(m_D+m_S + \Delta m\big).
\end{align}
The mass degeneracy between $\chi_2^0$ and $\chi^\pm$ at tree level is lifted by quantum corrections involving virtual gauge bosons. We have calculated these corrections and found them to lift the charged state above the neutral state by a few hundred MeV, so that $m_2\lesssim m_c$. In our numerical analysis of anomalous Higgs couplings, we neglect the small mass splitting between $\chi_2^0$ and $\chi^\pm$.

In this work, we will focus on scenarios with strong, but still perturbative dark Yukawa couplings,
\begin{align}
1 \lesssim y \lesssim \sqrt{4\pi}.
\end{align}
This parameter region is different from the bino-higgsino scenario in the MSSM, where the neutralino mixing is determined by electroweak interactions. The phenomenology of our model will thus look drastically distinct from the widely investigated scenarios with small Higgs couplings to dark fermions. In particular, large dark Yukawa couplings imply a large mass splitting $\Delta m \gtrsim 2yv$ among the neutral fermions in the dark sector.

Besides Yukawa couplings, dark fermions with weak quantum numbers also have electroweak interactions with the SM. In terms of mass eigenstates, the Higgs and gauge couplings are given by
\begin{align}\label{eq:higgs-gauge}
{\cal L} \supset & \ i \frac{g}{2c_W} \big(\sin\theta\, \chi_3^{0\ast} - \cos\theta\, \chi_1^{0\ast}\big) \bar \sigma^\mu \chi_2^0 Z_\mu + {\rm h.c.} + \frac{g}{c_W}\Big[\chi^- \bar \sigma^\mu (\textstyle{\frac{1}{2}} - s_W^2)\chi^+ - \chi^+ \bar \sigma^\mu (\textstyle{\frac{1}{2}} - s_W^2)\chi^-\Big] Z_\mu\\\nonumber
& + \frac{g}{2}\Big[\chi^+\bar\sigma^\mu \big(\cos\theta\,\chi_1^0 - \sin\theta\,\chi_3^0\big) - \big(\cos\theta\,\chi_1^{0\ast} - \sin\theta\,\chi_3^{0\ast}\big)\bar\sigma^\mu\chi^+ + i \big(\chi_2^{0\ast}\bar \sigma^\mu \chi^+ - \chi^+\bar \sigma^\mu\chi_2^0\big) \Big] W^-_\mu\\\nonumber
& - \frac{g}{2}\Big[\chi^-\bar\sigma^\mu \big(\cos\theta\,\chi_1^0 - \sin\theta\,\chi_3^0\big) - \big(\cos\theta\,\chi_1^{0\ast} - \sin\theta\,\chi_3^{0\ast}\big)\bar\sigma^\mu\chi^- - i \big(\chi_2^{0\ast}\bar \sigma^\mu \chi^- - \chi^-\bar \sigma^\mu\chi_2^0\big) \Big] W^+_\mu\\\nonumber
&  + e\chi^+\bar \sigma^\mu\chi^-A_\mu - \frac{y}{2} \Big[\sin(2\theta)\big(\chi_3^{0}\chi_3^0 - \chi_1^{0}\chi_1^0\big) - 2 \cos(2\theta)\chi_3^{0}\chi_1^0 \Big]h + {\rm h.c.}.
\end{align}
Dark fermion interactions with the $Z$ boson are off-diagonal in the mass basis, as a consequence of their Majorana nature. For the phenomenology of the model, especially for its interpretation in terms of dark matter, it is convenient to distinguish between two different scenarios,
\begin{align}\label{eq:scenarios}
1)\ |m_1| < m_2\lesssim m_c < m_3: & \ \text{ lightest state }  \chi_1^0,\\\nonumber
2)\ m_2 \lesssim m_c < |m_1| < m_3: & \ \text{ lightest state } \chi_2^0.
\end{align}
In scenario 1, the lightest state $\chi_1^0$ is a mixture of weak doublet and singlet components that couples to the Higgs boson. In scenario 2, the lightest state $\chi_2^0$ is a pure weak doublet that couples neither to the Higgs, nor to the $Z$ boson.

\begin{figure}[t]
\centering
\includegraphics[height=2.5in]{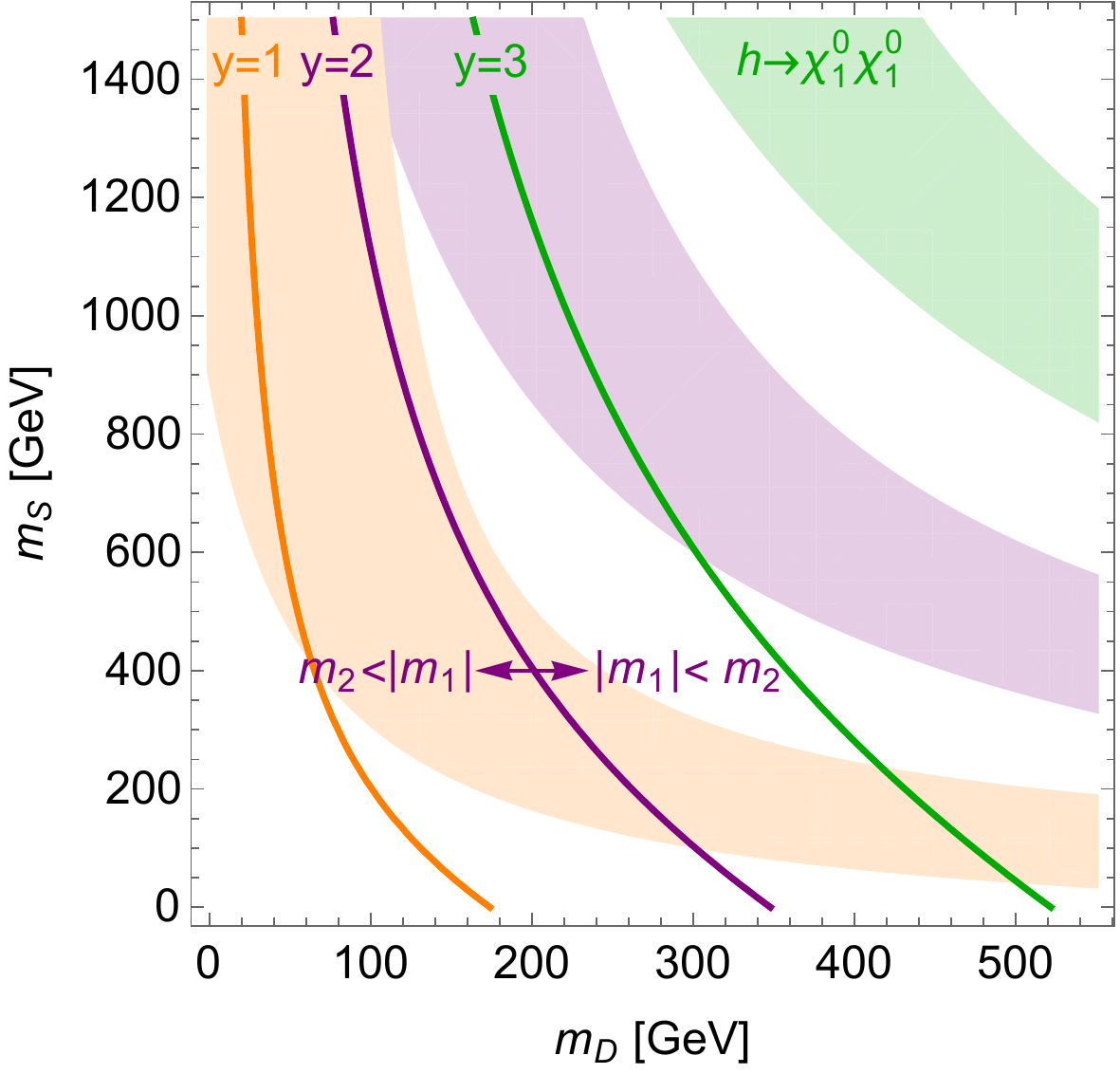} \hspace*{0.5cm} 
\caption{\label{fig:dm}Parameter space of possible lightest neutral states $\chi_1^0$ and $\chi_2^0$ in the dark fermion model. Colored lines separate regions with $|m_1| < m_2$ (right) and $m_2 < |m_1|$ (left) for fixed values of $y=1,2,3$. Colored areas are excluded by bounds on invisible Higgs decays $h\to \chi_1^0\chi_1^0$.}
\end{figure}
In Fig.~\ref{fig:dm}, we display the parameter space of our model in terms of the mass parameters $m_D$ and $m_S$, for fixed dark Yukawa couplings $y=1$ (orange), $y=2$ (purple), and $y=3$ (green), respectively. The colored curves separate the two scenarios of fermion mass spectra. To the right of each curve, scenario 1 is realized with $|m_1| < m_2$. To the left, one has $m_2 < |m_1|$, which corresponds to scenario 2. The latter scenario is thus favored for small values of $m_D$ and a large mass splitting $\Delta m$ among the neutral states. In the colored regions, the mass of $\chi_1^0$ is less than half the Higgs mass, $|m_1| < M_h/2$, so that the Higgs boson can decay invisibly via $h\to \chi_1^0\chi_1^0$. Current measurements of invisible Higgs decays at the LHC set a limit on the branching ratio~\cite{Aad:2015pla,Khachatryan:2016whc},
\begin{align}
\mathcal{B}(h\to \chi_1^0\chi_1^0) \lesssim 0.25,
\end{align}
which excludes dark fermion masses $|m_1| < M_h/2$. In the region $m_2 < M_h/2 < |m_1|$, invisible Higgs decays do not occur at appreciable rates, since the decay $h\to \chi_2^0\chi_2^0$ is absent at tree level.

Interpreted as dark matter candidates, the lightest states in scenarios 1 and 2 from Eq.~(\ref{eq:scenarios}) have a very different phenomenology. In scenario 1, the dark matter candidate $\chi_1^0$ has couplings to the Higgs boson, which induces spin-independent dark matter-nucleon scattering. The absence of such a signal at direct detection experiments sets an extremely strict bound on the dark Yukawa coupling~\cite{Freitas:2015hsa}. In this scenario, effects in Higgs observables are thus not compatible with the results of direct dark matter detection experiments. In scenario 2, in turn, the lightest state $\chi_2^0$ does not couple diagonally to the Higgs and $Z$ bosons. Dark matter-nucleon scattering is only induced at the loop level through electroweak interactions. Scenario 2 is thus much better protected from direct detection bounds than scenario 1. For $\chi_2^0$ to be a thermal relic, strong co-annihilation with the nearly-degenerate charged states $\chi^\pm$ requires the mass spectrum of dark fermions to be around the TeV scale~\cite{Freitas:2015hsa,Xiang:2017yfs}. Alternatives to thermal freeze-out that lead to the observed dark matter relic density with a lighter spectrum are of course a possibility.

\section{Virtual dark fermions in Higgs interactions}\label{sec:higgs-df}
\noindent Due to the large dark Yukawa coupling, the dominant effects of dark fermions in collider observables are a priori expected to occur in Higgs interactions. A first idea that might come to mind is resonant production of dark fermion pairs through $pp\to h^\ast \to \chi_i\chi_j$. However,  in this process the Higgs boson is produced off-shell for fermion pair invariant masses above the Higgs mass. The production rate is thus suppressed by the small Higgs decay width and hard to observe at the LHC. Similar complications arise for electroweak production via off-shell gauge bosons, $pp\to W^\ast/Z^\ast \to \chi_i\chi_j$. More details of possible direct collider searches will be discussed in Section~\ref{sec:resonant}.

\begin{figure}[t]
\centering
\includegraphics[height=1in]{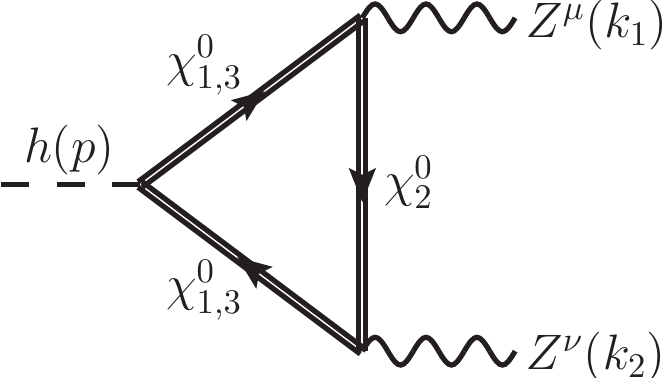} \hspace*{1cm} \includegraphics[height=1in]{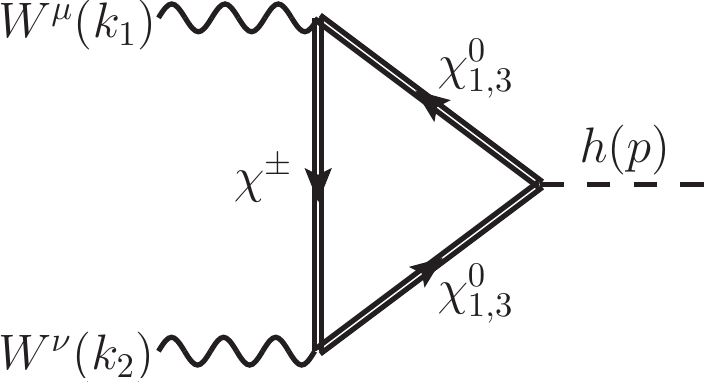} \hspace*{1cm} \includegraphics[height=1in]{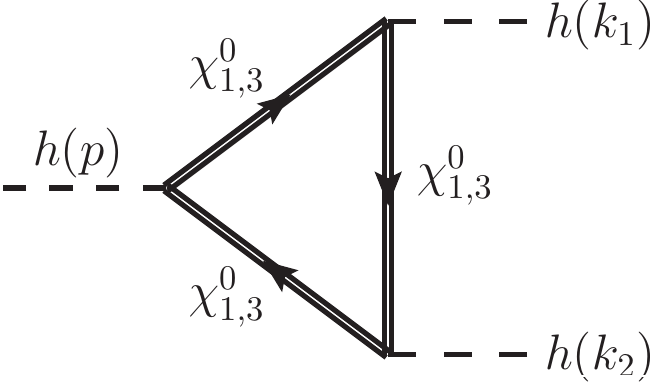}
\caption{\label{fig:hvv-hhh}Feynman diagrams for vertex corrections from dark fermions to $h\to ZZ$ (left), $WW\to h$ (center), and $h\to hh$ (right).}
\end{figure}

Here we argue that dark fermions with large Yukawa couplings can be probed \emph{indirectly} through virtual effects in Higgs couplings to weak gauge bosons and Higgs self-interactions. Examples of Feynman diagrams are shown in Fig.~\ref{fig:hvv-hhh}. Higgs self-interactions receive the largest corrections, due to their strong sensitivity to the dark Yukawa coupling. Among Higgs-gauge boson couplings, $h\gamma\gamma$ and $hZ\gamma$ interactions are expected to be most sensitive to new virtual corrections, since these couplings are loop-suppressed in the SM. Contributions from new fermions, however, require a renormalizable Higgs coupling to two charged states with different weak quantum numbers. In minimal models such as ours, $h\gamma\gamma$ and $hZ\gamma$ are not affected by dark fermions at the one-loop level. The main modifications of Higgs-gauge boson interactions occur in $hZZ$ and $hWW$ couplings. Since these couplings arise as fundamental interactions from the Higgs kinetic terms in the SM, relative corrections from dark fermions are expected to be modest. As we will show, in the regime of strong Yukawa couplings they are sizable enough to be probed at the LHC and even better so at a future lepton collider.

In this section, we will systematically analyze the main sub-processes that probe $hVV$ ($V=W,Z$) and $hhh$ interactions at the LHC and (with certain modifications) at future lepton colliders. For Higgs-gauge interactions, these are Higgs decays to gauge boson pairs, $h\to VV^\ast$; weak boson fusion, $V^\ast V^\ast \to h$; and gauge-boson associated Higgs production, $V^\ast \to Vh$. Each of these sub-processes probes the anomalous $hVV$ interaction in a different kinematic region, which makes them a priori complementary indirect searches for dark fermions. Triple Higgs interactions can be directly probed in Higgs pair production, $h^\ast\to hh$.

\subsection{Higgs-gauge boson interactions}\label{sec:hvv}
\noindent Consider the decay of a resonantly produced Higgs boson into a pair of weak gauge bosons, $h(p)\to V(k_1)V^\ast(k_2)$, as shown for $V=Z$ in Fig.~\ref{fig:hvv-hhh}, left. Here $p$ is the four-momentum of the incoming Higgs, and $k_1$, $k_2$ are the four-momenta of the outgoing vector bosons. Momentum conservation at the vertex implies $p^\mu = k_1^\mu + k_2^\mu$. At colliders, this sub-process is observed through a four-fermion final state obtained from the gauge boson decays, $VV^\ast\to (f\bar{f})(f\bar{f})$. We work in the approximation that one vector boson, $V(k_1)$, is produced on its mass shell and neglect light fermion masses. In $CP$-conserving theories, virtual corrections from new particles can then generally be expressed in terms of two kinematic form factors, $F_V^0(p^2, k_1^2, k_2^2)$ and $F_V^1(p^2, k_1^2, k_2^2)$, as~\cite{Choi:2002jk}
\begin{align}
\delta \Gamma_V^{\mu\nu}(h(p)\to V^\mu(k_1)V^\nu(k_2)) & \equiv g_{hVV}\Big[F_V^0(p^2,k_1^2, k_2^2)\,g^{\mu\nu} + F_V^1(p^2, k_1^2, k_2^2)\,\frac{k_{2}^{\mu}k_{1}^{\nu}}{M_V^2}\Big].
\end{align}
Here $g_{hVV}$ denotes either of the tree-level SM couplings $g_{hZZ} = e M_Z/(s_W c_W)$ or $g_{hWW} = e M_W/s_W$. The vertex contribution to $F_V^0$, denoted as $F_V^{0,\text{bare}}$ in what follows, is in general UV-divergent and needs to be renormalized by a counter term $\delta F_V^0$. In this work, we employ the on-shell renormalization scheme, following the notation from Ref.~\cite{Denner:1991kt}. The finite renormalized form factor is given by
\begin{align}\label{eq:counterv}
F_V^0(p^2, k_1^2, k_2^2) & = F_V^{0,\text{bare}}(p^2, k_1^2, k_2^2) + \delta F_V^0,\\\nonumber
\delta F_V^0 & = \delta Z_e + \Big(\frac{2s_W^2}{c_W^2} \delta_{VZ} - 1\Big)\frac{\delta s_W}{s_W} + \frac{1}{2}\frac{\delta M_W^2}{M_W^2} + \frac{1}{2}\delta Z_H + \delta Z_V,
\end{align}
where $\delta_{VZ}=1(0)$ if $V$ is the $Z(W)$ boson. The vertex correction $F_V^1$ is finite. In our model, the virtual contributions to $h\to VV^\ast$ from the dark fermions in their mass eigenstates are
\begin{align}\label{eq:hvv-vertex}
g_{hVV} F_V^{0,\text{bare}}(p^2, k_1^2, k_2^2) & = \frac{-i}{8\pi^2}\sum_{ij=11,13,31,33} g^{ij}_h\,g^{jD}_V\,g^{Di}_V\,L_V^0(p^2, k_1^2, k_2^2, m_i,m_j,m_D),\\\nonumber
g_{hVV} F_V^1(p^2, k_1^2, k_2^2) & = \frac{-i}{8\pi^2}\sum_{ij=11,13,31,33} g^{ij}_h\,g^{jD}_V\,g^{Di}_V\,L_V^1(p^2, k_1^2, k_2^2, m_i,m_j,m_D),
\end{align}
where the index $D=2$ for $V=Z$ and $D=c$ for $V=W$. The Higgs and gauge-boson couplings of the dark fermions, $g_h$ and $g_V$, as well as the loop functions, $L_V^0$ and $L_V^1$, are given in Appendix~\ref{app:hvv}. Explicit expressions for the counter terms in Eq.~(\ref{eq:counterv}) can be obtained from Appendix~\ref{app:cts}.\\

As our second probe of dark fermions, we consider Higgs production via weak boson fusion, $V^\ast(k_1)V^\ast(k_2)\to h(p)$, where the four-momenta $k_1$, $k_2$ are defined as incoming and $p=k_1+k_2$ as outgoing. Since the relative orientation of external momenta in weak boson fusion is the same as in Higgs decays, the structure of the vertex correction in both processes is the same,
\begin{align}\label{eq:vvh-vertex}
\delta\Gamma_V^{\mu\nu}(V^{\mu}(k_1)V^{\nu}(k_2)\to h(p)) & = g_{hVV}\Big[F_V^0(p^2,k_1^2, k_2^2)\,g^{\mu\nu} + F_V^1(p^2, k_1^2, k_2^2)\,\frac{k_{2}^{\mu}k_{1}^{\nu}}{M_V^2}\Big]\\\nonumber
& = \delta\Gamma_V^{\mu\nu}(h(p) \to V^{\mu}(k_1)V^{\nu}(k_2)).
\end{align}
As in Higgs decays, we have neglected the masses of light fermions coupling to the gauge bosons in weak boson fusion. Notice that the momenta $k_1^2$ and $k_2^2$ that are probed in weak boson fusion and Higgs decays are different (see Section~\ref{sec:formfactors}).\\

The third sub-process of anomalous Higgs-gauge interactions is the associated resonant production of a Higgs and a gauge boson via $V^\ast(k_1)\to V(k_2) h(p)$. Since $k_1$ is an incoming four-vector, whereas $k_2$ and $p=k_1-k_2$ are outgoing momenta, the momentum constellation is different from Higgs decays and weak boson fusion. The one-loop correction to associated production can be expressed as
\begin{align}\label{eq:vhv-vertex}
\delta\Gamma_V^{\mu\nu}(V^{\mu}(k_1)\to V^{\nu}(k_2)h(p)) & = g_{hVV}\Big[F_V^0(p^2,k_1^2,k_2^2)\,g^{\mu\nu} + \widetilde{F}_V^1(p^2,k_1^2,k_2^2)\frac{k_2^{\mu}k_1^{\nu}}{M_V^2} \Big],\\\nonumber
g_{hVV} \widetilde{F}_V^1(p^2, k_1^2, k_2^2) & = \frac{-i}{8\pi^2}\sum_{ij=11,13,31,33} g^{ij}_h\,g^{jD}_V\,g^{Di}_V\,\widetilde{L}_V^1(p^2, k_1^2, k_2^2, m_i,m_j,m_D).
\end{align}
The first form factor $F_V^0$ is the same in all three processes. The second form factor $\widetilde{F}_V^1$ is sensitive to the momentum configuration of the gauge bosons and thus differs from $F_V^1$ in Higgs decays and weak boson fusion. An explicit expression of the corresponding loop function $\widetilde{L}_V^1$ in our model is given in Appendix~\ref{app:hvv}.

\subsection{Triple Higgs interactions}\label{sec:hhh}
\noindent Higgs self-interactions can be directly probed through the sub-process $h^\ast(p)\to h(k_1)h(k_2)$, where $p=k_1+k_2$ is incoming and $k_1$, $k_2$ are outgoing four-momenta. The one-loop corrections to this process from dark fermions can be written as
\begin{align}
\delta \Gamma_h(h(p)\to h(k_1) h(k_2)) \equiv \lambda_{3} F_h(p^2, k_1^2, k_2^2),
\end{align}
with the tree-level triple Higgs coupling in the SM, $\lambda_{3} = 3M_h^2/v$. The scalar form factor $F_h(p^2, k_1^2, k_2^2)$ is obtained after renormalization as~\cite{Denner:1991kt}
\begin{align}\label{eq:counterh}
F_h(p^2, k_1^2, k_2^2) & = F_h^{\text{bare}}(p^2, k_1^2, k_2^2) + \delta F_h,\\\nonumber
\delta F_h & = \delta Z_e - \frac{\delta s_W}{s_W} + \frac{\delta M_h^2}{M_h^2} + \frac{e}{2s_W}\frac{\delta t}{M_W M_h^2} - \frac{1}{2}\frac{\delta M_W^2}{M_W^2} + \frac{3}{2}\delta Z_h.
\end{align}
In our model, the vertex correction induced by virtual dark fermions can be expressed as
\begin{align}
\lambda_{3} F_h^{\text{bare}}(p^2, k_1^2, k_2^2) & = \frac{i}{8\pi^2}\Big[\sum_{ij=11,13,31,33} g^{ij}_h\,g^{ji}_h\,g^{ii}_h\,L_h(p^2, k_1^2, k_2^2, m_i,m_j)\\\nonumber
& \hspace*{1.8cm}+ \sum_{ij=13,31} g^{ij}_h\,g^{ji}_h\,g^{ii}_h\,\Big\{L_h(k_2^2, p^2, k_1^2, m_i,m_j) + L_h(k_1^2, k_2^2, p^2, m_i,m_j)\Big\}\Big].
\end{align} 
Explicit expressions for the loop function $L_h$ and the counter terms in Eq.~(\ref{eq:counterh}) can be found in Appendices~\ref{app:hvv} and \ref{app:cts}, respectively.

\section{Higgs phenomenology at colliders}\label{sec:formfactors}
\noindent In the previous section, we have derived the structure of dark fermion contributions to anomalous Higgs interactions in terms of form factors. We now analyze these form factors numerically in the kinematic regions that are relevant for collider observables. Details of the observation prospects at the LHC and future lepton colliders will be discussed in Section~\ref{sec:future}. In processes with resonant Higgs production, we apply the narrow-width approximation, $\Gamma_h\to 0$, so that Higgs production and decay factorize. In Higgs decay, $h\to VV^\ast$, and Higgs-gauge boson associated production, $V^\ast\to Vh$, we furthermore assume one of the gauge bosons to be on-shell.\\

Let us first analyze the Higgs couplings to gauge bosons. At the LHC, the sub-processes described in Section~\ref{sec:hvv} can be probed in four hadronic processes: 1) inclusive Higgs production and decay; 2) Higgs production through weak boson fusion; 3) associated Higgs-gauge boson production; 4) top-pair associated Higgs production and decay. These processes are defined in the following kinematic regions,
\begin{align}
1)\ & pp\to h(p^2)\to V(k_1) V^\ast(k_2): \quad k_1^2 = M_V^2,\ 0 \leq k_2^2 \leq (M_h-M_V)^2,\\\nonumber
2)\ & pp\to V^\ast(k_1) V^\ast(k_2) j_1 j_2 \to h(p^2) j_1 j_2: \quad k_1^2, k_2^2 < 0,\\\nonumber
3)\ & pp\to V^\ast(k_1)\to V(k_2)h(p^2): \quad k_1^2 \gtrsim (M_h+M_V)^2,\ k_2^2 = M_V^2,\\\nonumber
4)\ & pp\to t\bar t\,h(p^2) \to t\bar t\,V(k_1) V^\ast(k_2): \quad k_1^2 = M_V^2,\ 0 \leq k_2^2 \leq (M_h-M_V)^2.
\end{align}
In all processes, the Higgs boson is assumed to be on-shell, $p^2=M_h^2$. Notice that the sub-process $h\to VV^\ast$ can also be probed in Higgs production through weak boson fusion or Higgs-gauge associated production with subsequent Higgs decay into gauge boson pairs. In Higgs decays and weak boson fusion, the dependence of the form factor $F_V^0(M_h^2,k_1^2,k_2^2)$ on the squared momenta of the off-shell boson(s) is very weak. A good numerical estimate of the effect on the observables can thus be obtained by considering the kinematic reference value
\begin{align}\label{eq:ref-hvv}
h\to VV^\ast,V^\ast V^\ast \to h\ (pp): & \qquad F_V^{0,pp}\equiv F_V^0(M_h^2,M_V^2,(25\,\text{GeV})^2).
\end{align}
The reference value $k_2^2 = (25\,\text{GeV})^2$ corresponds with the maximum of the distribution $d\Gamma(h\to VV^\ast \to (f\bar f) (f\bar f))/d M_{f\bar f}$, where $M_{f\bar f}^2 = k_2^2$ is the invariant mass of the decay products of the off-shell boson. In Higgs-gauge associated production, the momentum of the intermediate vector boson is determined by the partonic center-of-mass energy.

At electron-positron colliders, associated Higgs-$Z$ production via $e^+ e^-\to Z^\ast \to Zh$~\cite{Ellis:1975ap,Ioffe:1976sd,Glashow:1978ab} and Higgs production through $W$ fusion $e^+ e^-\to h\nu\bar{\nu}$~\cite{Jones:1979bq,Kilian:1995tr} are the most important processes to measure Higgs-$Z$ and Higgs-$W$ interactions, respectively. In $e^+ e^-\to Z^\ast \to Zh$, the virtuality of the off-shell $Z$ boson is set by the collider energy,
\begin{align}
e^+ e^- \to Z^\ast(k_1)\to Z(k_2)h(p): & \quad k_1^2 = s,\ k_2^2 = M_Z^2,\ p^2=M_h^2.
\end{align}
As our reference point, we choose a scenario planned for the ILC with $\sqrt{s}=250\,\text{GeV}$, so that
\begin{align}\label{eq:ref-zzh}
Z^\ast \to Zh\ (e^+e^-): \quad F_Z^{0,ee} \equiv F_Z^0(M_h^2,(250\,\text{GeV})^2,M_Z^2).
\end{align}
The magnitude of the form factor decreases with increasing $\sqrt{s}$. Measuring associated Higgs-$Z$ boson production in $e^+e^-$ collisions at different energies can therefore be used to probe the momentum dependence of $F_Z^0(M_h^2,k_1^2,M_Z^2)$.

So far, we have focused our attention on the form factor $F_V^0$, which rescales the SM coupling $g_{hVV}$ by a momentum-dependent factor. The form factors $F_V^1$ and $\widetilde{F}_V^1$ introduce new effective interactions (see Eqs.~(\ref{eq:vvh-vertex}) and (\ref{eq:vhv-vertex})), which have no equivalent in the SM and could change the kinematics of the respective process. In our case, however, it turns out that $F_V^1$ and $\widetilde{F}_V^1$ are numerically tiny. In the phase space and parameter space relevant for collider phenomenology, they are more than two orders of magnitude smaller than $F_V^0$. We therefore discard $F_V^1$ and $\widetilde{F}_V^1$ from our analysis in what follows. Notice furthermore that the only difference between $F_W^0$ and $F_Z^0$ are small phase-space effects due to the different $W$ and $Z$ boson masses. The reason is that $hWW$ vertex corrections can be obtained from $hZZ$ corrections by replacing $\chi_2^0$ with $\chi^\pm$ (see Fig.~\ref{fig:hvv-hhh} and Eq.~(\ref{eq:higgs-gauge})), which are nearly mass-degenerate. In our phenomenological analysis, we will focus on Higgs observables with $Z$ bosons, for which often a better experimental precision is expected. Using the fact that $F_W^0\approx F_Z^0$, our results can easily be translated to processes with $W$ bosons.

\begin{figure}[t]
\centering
\includegraphics[height=2.1in]{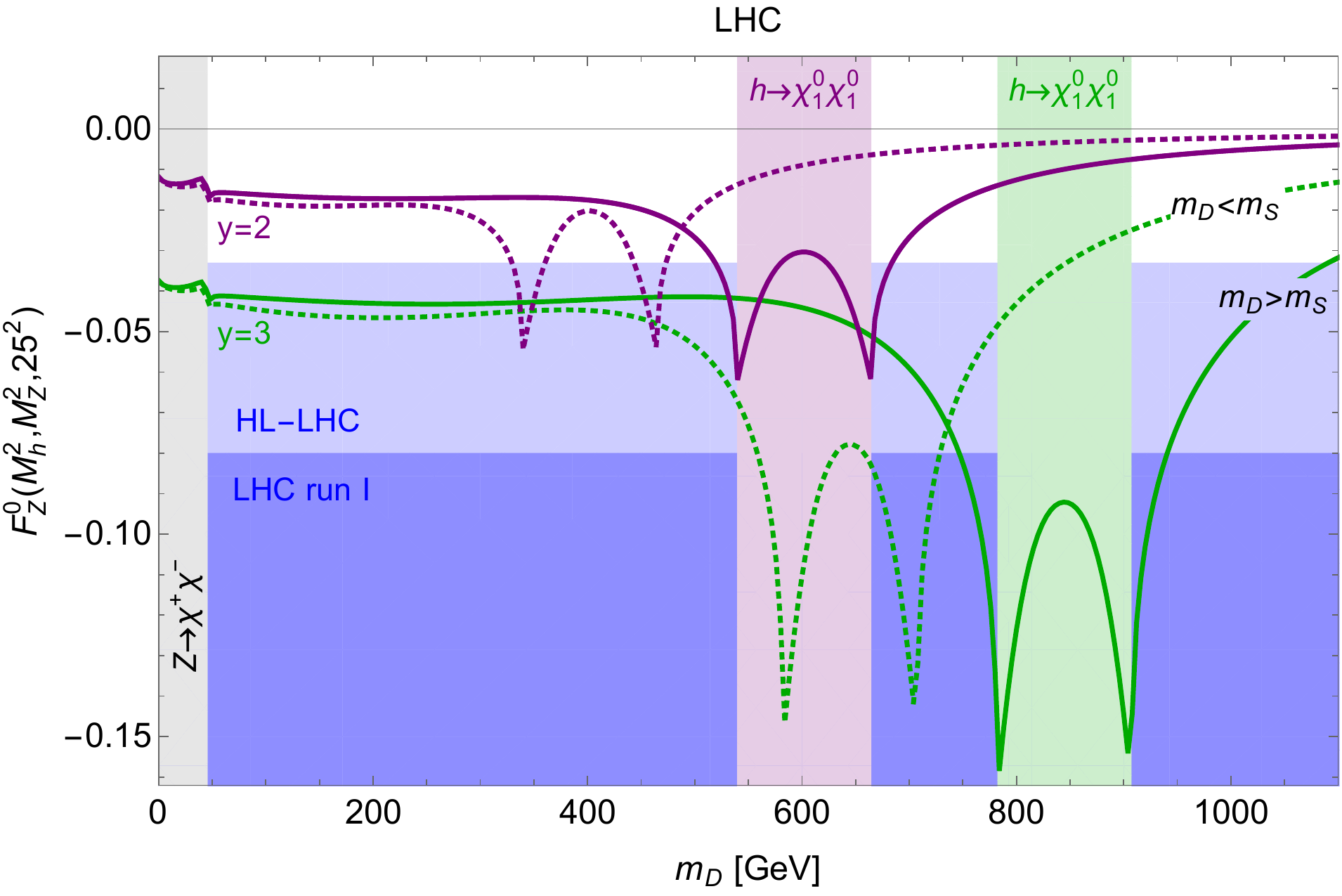} \hspace*{0.1cm} \includegraphics[height=2.1in]{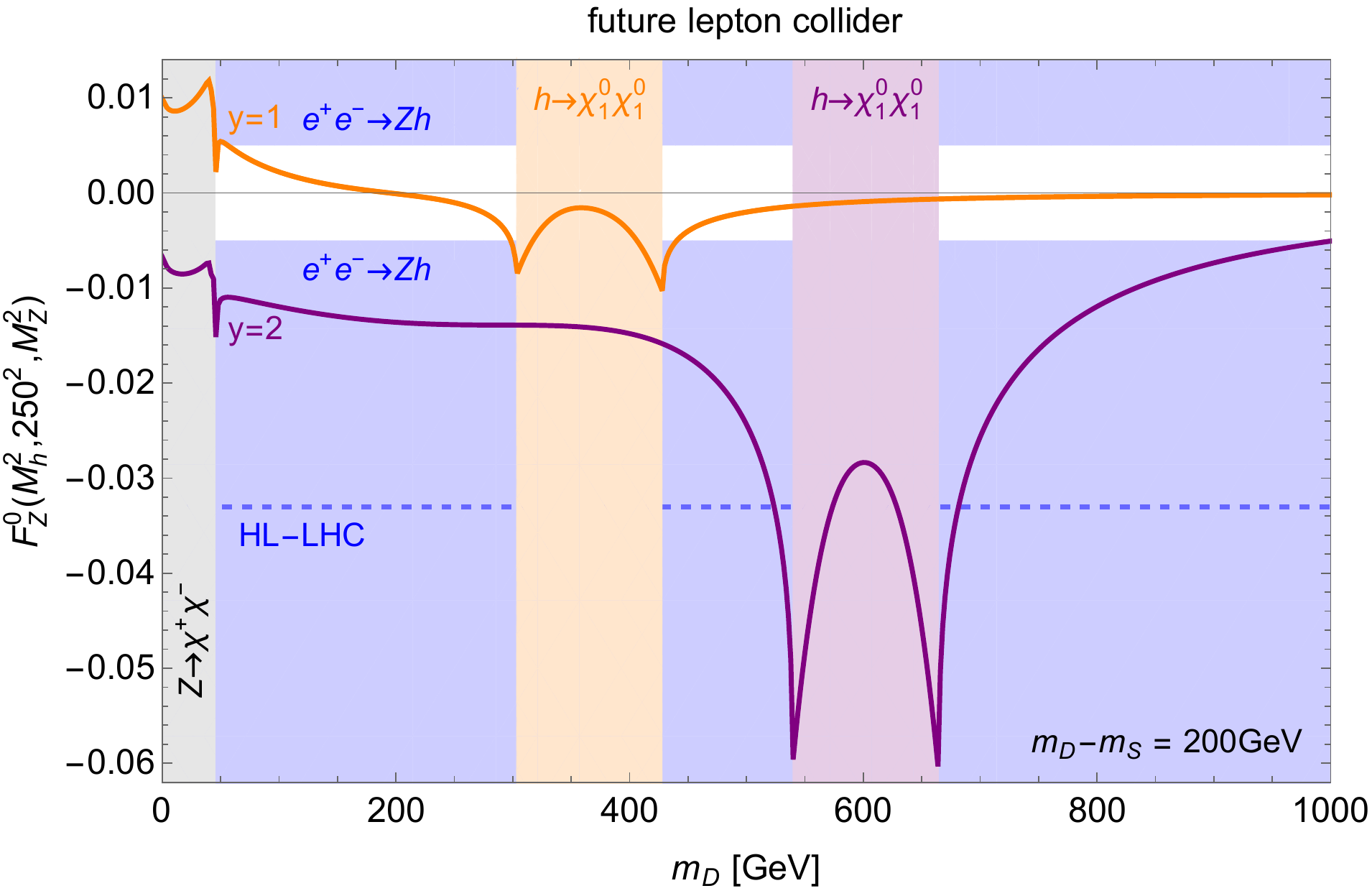}
\caption{\label{fig:f0-parameters}Anomalous Higgs-$Z$ coupling as a function of the mass parameter $m_D$ for fixed Yukawa couplings $y$ at the LHC, $\text{Re}[F_Z^{0,pp}]$, (left) and a future lepton collider with $\sqrt{s}=250\,\text{GeV}$, $\text{Re}[F_Z^{0,ee}]$, (right). Plain/dotted lines correspond to fixed values $m_D - m_S = \pm200\,\text{GeV}$. The gray area is excluded by $Z$ width measurements. In orange, purple, and green regions, invisible Higgs decay $h\to \chi_1^0\chi_1^0$ is open for $m_D - m_S = 200\,\text{GeV}$ and $y=1,2,3$, respectively. Dark blue regions have been excluded at the LHC during run I. Light blue regions can be tested at the HL-LHC and future lepton colliders.}
\end{figure}
Having assessed the kinematic features of the form factors, we now study their dependence on the model parameters $m_D$, $m_S$ and $y$. In Fig.~\ref{fig:f0-parameters}, we show the real part of $F_Z^0$ at the kinematic reference points relevant for Higgs decays to $Z$ boson pairs at the LHC (left), and for associated Higgs-$Z$ production at a future lepton collider with $\sqrt{s}=250\,\text{GeV}$ (right). For illustration, we have fixed the dark Yukawa coupling to $y=1$ (orange), $y=2$ (purple) and $y=3$ (green), as well as the mass parameter difference $m_D - m_S = \pm 200\,\text{GeV}$ (plain/dotted curves).\footnote{For other mass parameter splittings, the features of the form factors are qualitatively very similar.} The doublet mass $m_D$ remains a free parameter. Parameter regions that are excluded by invisible Higgs decays for $m_D-m_S = 200\,\text{GeV}$ are displayed as colored areas. The gray area, where $m_D < M_Z/2$, is excluded by measurements of the $Z$ width, which would be enlarged by decays into charged dark fermions, $Z\to \chi^+\chi^-$~\cite{ALEPH:2005ab}. The dark blue region has been excluded by a global analysis of Higgs couplings with LHC data from run I~\cite{Corbett:2015ksa}.

The masses of the charged and neutral states $\chi^\pm$ and $\chi_2^0$, $m_c = m_D = m_2$, can be directly read off from the $x$-axis. For $m_D < M_Z/2$, the state $\chi_2^0$ in the loop can be on its mass shell and the form factor $F_Z^0$ develops an imaginary part. This explains the peak-dip feature of the real part in the region around $m_D\approx 45\,\text{GeV}$. For $m_D\to \infty$, effects of the dark sector decouple from the SM. The mass parameter difference $|m_D - m_S|$ and the Yukawa coupling $y$ determine the splitting $\Delta m$ between the lightest and heaviest states $\chi_1^0$ and $\chi_3^0$ (see Eq.~(\ref{eq:masses})). As we can observe in the figure, sizable effects of dark fermions on the Higgs interactions require a large Yukawa coupling. In scenario 1, this implies a split spectrum with a light state $\chi_1^0$, intermediate states $\chi_2^0,\,\chi^\pm$, and a heavy state $\chi_3^0$. Notice that $F_Z^0$ is largest close to the parameter regions excluded by $h\to \chi_1^0\chi_1^0$, where $|m_1| \gtrsim M_h/2$. In this region, $F_Z^0$ is dominated by the loop diagram in Fig.~\ref{fig:hvv-hhh}, left, with two lightest states $\chi_1^0$ and one $\chi_2^0$.

The effects of dark fermions in Higgs decays can be directly translated to weak boson fusion at the LHC. Due to the small momentum dependence of $F_Z^0$ in these processes, virtual corrections in $Z^\ast Z^\ast \to h$ look almost identical to those displayed in Fig.~\ref{fig:f0-parameters}, left. Effects in Higgs-$Z$ associated production at the LHC are comparable in size with the FLC reference point, $F_Z^0(M_h^2,k_1^2\gtrsim (M_h + M_Z)^2,M_Z^2)\approx F_Z^{0,ee}$, shown in Fig.~\ref{fig:f0-parameters}, right. Comparing Higgs decays (and likewise Higgs production from $Z$ boson fusion) at the LHC with Higgs-$Z$ associated production at a FLC (and likewise at the LHC), we observe that $F_Z^{0,ee}$ for associated production is slightly smaller than $F_Z^{0,pp}$ for decay and weak fusion. However, the expected precision of measuring $F_Z^0$ at a lepton collider is much higher than at the LHC. The light blue areas in Fig.~\ref{fig:f0-parameters} are expected to be probed at the HL-LHC with $3\,\text{ab}^{-1}$ data luminosity (left) and at a FLC with $\sqrt{s}=250\,\text{GeV}$ (right).\\

Higgs self-interactions can be analyzed in a similar way. At the LHC, the triple Higgs coupling can be directly measured in Higgs pair production, based on the sub-process $h^\ast \to hh$ from Section~\ref{sec:hhh}. The kinematic region for this process is given by
\begin{align}
pp\to h^\ast(p) \to h(k_1)h(k_2): \quad k_1^2 = M_h^2=k_2^2,\ p^2 \gtrsim (2M_h)^2.
\end{align}
Since the cross section of Higgs pair production drops quickly for higher invariant mass, we choose our kinematic reference point of the form factor near the production threshold,
\begin{align}\label{eq:ref-hhh}
h^\ast \to hh\ (pp,e^+e^-): \quad \overline{F}_h \equiv F_h((280\,\text{GeV})^2,M_h^2,M_h^2).
\end{align}
At a future lepton collider with energy $\sqrt{s} > 2M_h + M_Z$, the triple Higgs coupling can be measured in Higgs pair production in association with a $Z$ boson~\cite{Djouadi:1999gv},
\begin{align}
e^+e^-\to Z^\ast \to Zh^\ast(p) \to Zh(k_1)h(k_2): & \quad k_1^2 = M_h^2=k_2^2,\ (2M_h)^2 \le p^2 \le (\sqrt{s} - M_Z)^2.
\end{align}
This process is also sensitive to anomalous Higgs-$Z$ boson couplings. As we will see, in our model $F_Z^0$ is numerically much smaller than $F_h$. The process $e^+e^-\to Zhh$ can thus be considered as a clean probe of $F_h$. Since the cross section for $e^+e^-\to Zhh$ production is again dominated by Higgs-pair production near the kinematic threshold, the reference value $\overline{F}_h$ from Eq.~(\ref{eq:ref-hhh}) applies here as well.\footnote{The exact value of $F_h$ can vary with the momentum $p^2$, depending on the respective parameter point. However, since these variations are typically moderate and the cross section is largest near the production threshold, a good estimate of the overall effect can be obtained by studying the reference point $\overline{F}_h$.}

\begin{figure}[t]
\centering
\includegraphics[height=2.1in]{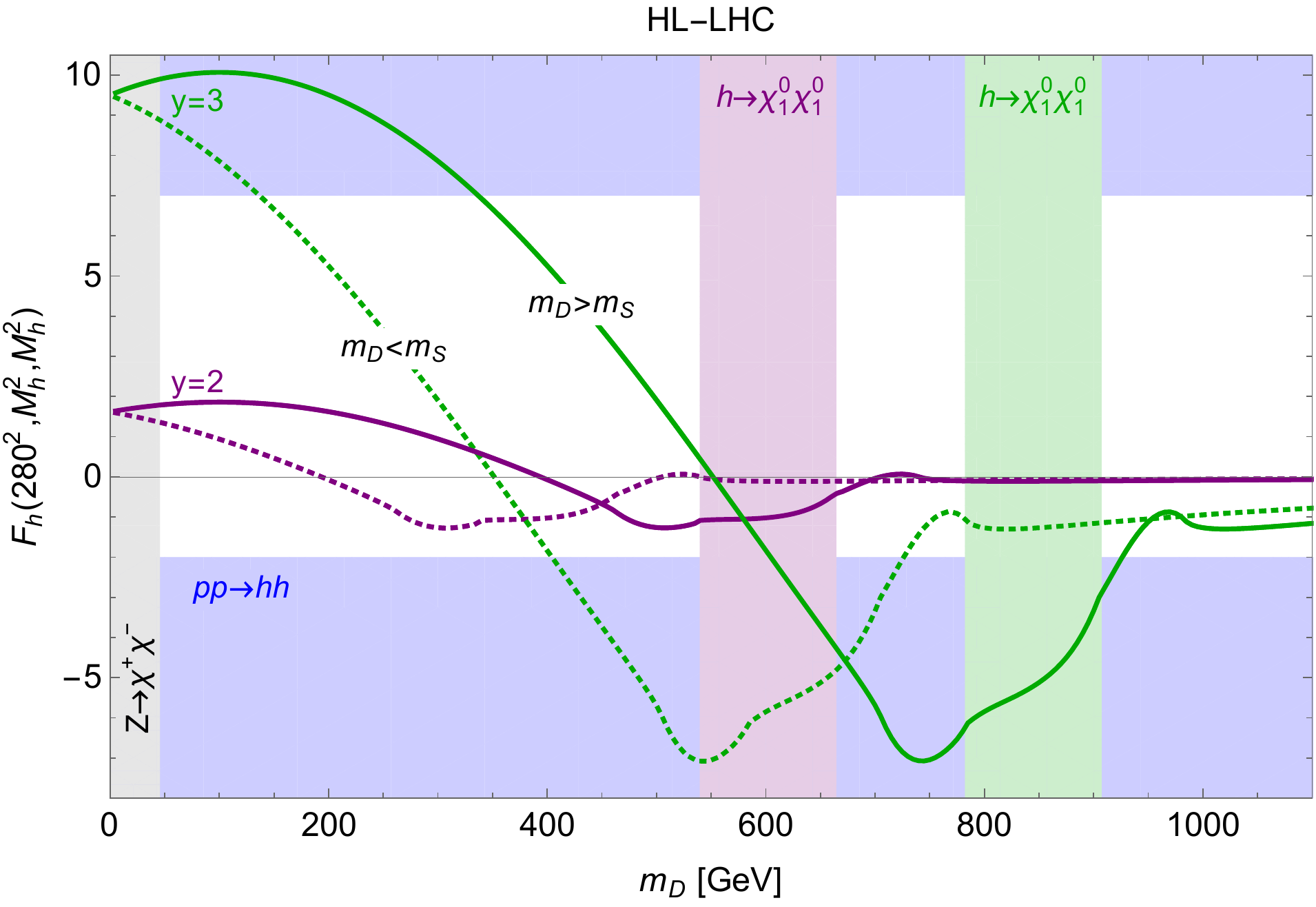} \hspace*{0.1cm} \includegraphics[height=2.1in]{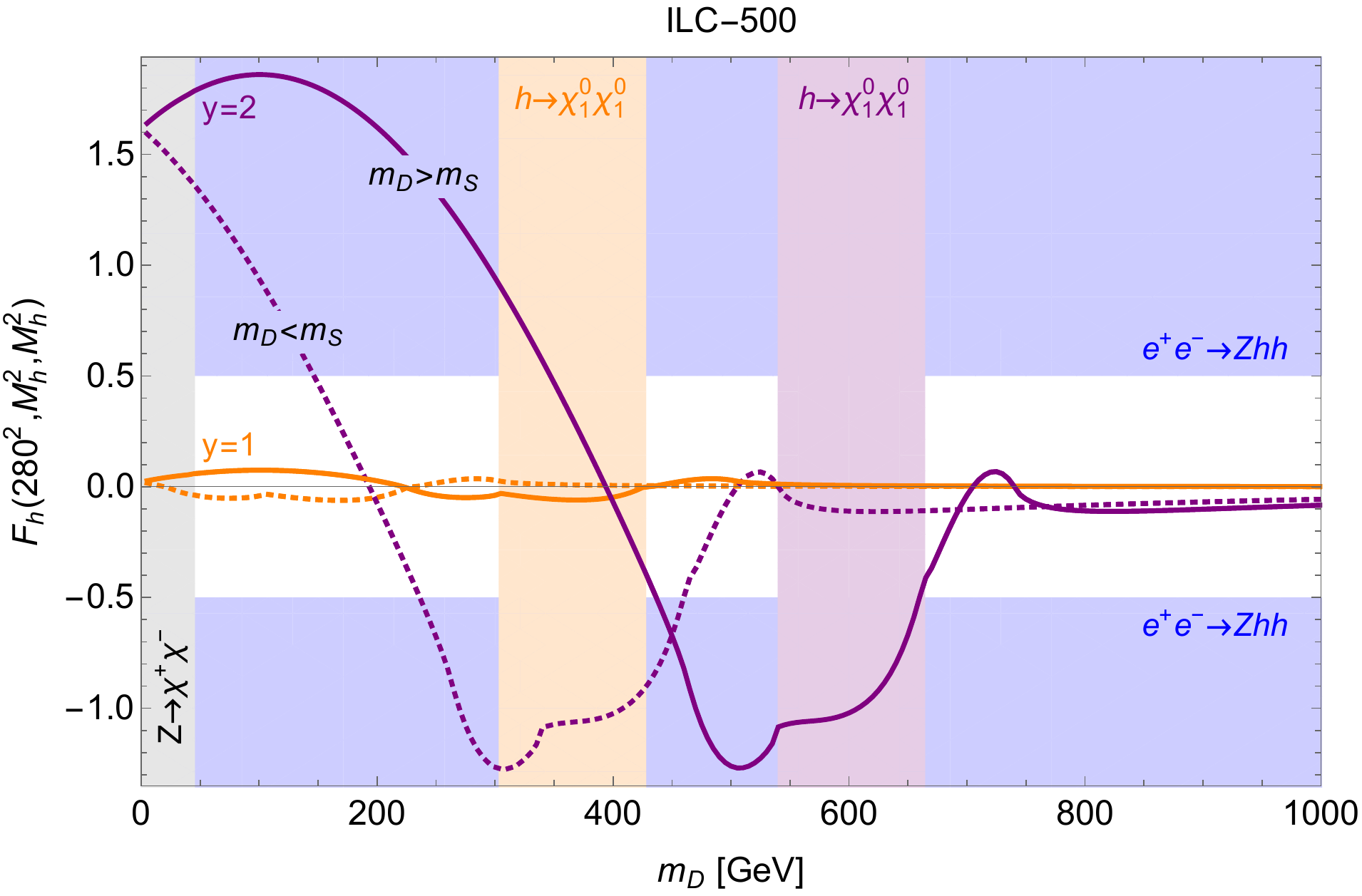}
\caption{\label{fig:f0h-parameters}Anomalous triple Higgs coupling, $\text{Re}[\overline{F}_h]$, as a function of the mass parameter $m_D$ for fixed Yukawa couplings $y$ at the HL-LHC (left) and a future ILC with $\sqrt{s}=500\,\text{GeV}$ (right). Plain/dotted lines correspond with $m_D - m_S = \pm 200\,\text{GeV}$. The gray area is excluded by $Z$ width measurements. In orange, purple, and green regions, invisible Higgs decay $h\to \chi_1^0\chi_1^0$ is open for $m_D - m_S = 200\,\text{GeV}$ and $y=1,2,3$, respectively. Blue regions can be probed at the HL-LHC and ILC-500, respectively.}
\end{figure}

In Fig.~\ref{fig:f0h-parameters}, we show the real part of $\overline{F}_h$ relevant for Higgs pair production at the HL-LHC (left) and $Z$-boson associated Higgs pair production at the ILC with $\sqrt{s} = 500\,\text{GeV}$ (right).
 Similarly to $F_Z^0$ in Fig.~\ref{fig:f0-parameters}, $F_h$ is large in the parameter region with $|m_1| \approx M_h/2$, where the loop function is dominated by the lightest state $\chi_1^0$ (see Fig.~\ref{fig:hvv-hhh}, right). The overall size of $F_h$, however, is one to two orders of magnitude larger than $F_Z^0$, due to the parametric dependence $F_h/F_Z^0 \sim y^2/g^2$. A second region of large $F_h$ is obtained for $|m_1|\approx m_3$. For $m_D - m_S = 200\,\text{GeV}$, exact mass degeneracy occurs at $m_D = 100\,\text{GeV}$. In this region of parameter space, both $\chi_1^0$ and $\chi_3^0$ contribute significantly to the vertex function. The blue areas are expected to be probed at the HL-LHC (left) and at the ILC with $\sqrt{s}=500\,\text{GeV}$ (right).
 
\section{Electroweak precision tests}\label{sec:ewpo}
\noindent Due to the electroweak couplings of dark fermions, contributions to electroweak precision observables are expected, which have been precisely measured at LEP. For new physics above the weak scale, such contributions can be analyzed in terms of the so-called oblique parameters $S$ and $T$, defined by~\cite{Peskin:1991sw}
\begin{align}
T & = \frac{4\pi}{e^2 c_W^2 M_Z^2}\Big[\Pi_{WW}(0) - c_W^2\Pi_{ZZ}(0) -2s_W c_W \Pi_{Z\gamma}(0) - s_W^2\Pi_{\gamma\gamma}(0) \Big],\\\nonumber
S & = \frac{16\pi s_W^2 c_W^2}{e^2}\Big[ \Pi_{ZZ}'(0) + \frac{s_W^2-c_W^2}{s_Wc_W}\Pi'_{Z\gamma}(0) - \Pi'_{\gamma\gamma}(0)\Big],
\end{align}
where $\Pi_{ij}(0)$ and $\Pi'_{ij}(0)$ denote contributions to the gauge boson two-point functions and their momentum derivative at zero momentum transfer, respectively. Deviations of $S$ and $T$ from their SM predictions due to new heavy particles are denoted by $\Delta S$ and $\Delta T$, respectively. Since the Lagrangian in Eq.~(\ref{eq:maj-sd}) preserves a custodial symmetry, the $T$ parameter is protected from contributions of dark fermions, resulting in $\Delta T=0$. Contributions to the $S$ parameter are moderate, since dark fermions have vector-like electroweak interactions. In the decoupling limit $m_S=m_D=m\gg v$, dark fermion contributions to the $S$ parameter in our model are given by
\begin{align}
\Delta S = \frac{1}{60\pi}\frac{y^2v^2}{m^2}\Big[1+\mathcal{O}\left(\frac{y^2v^2}{m^2}\right)\Big].
\end{align}
Precision measurements at LEP have set a limit on new physics contributions to $S$ at the $68\%$ C.L.~\cite{ALEPH:2005ab},
\begin{align}
 |\Delta S| < 0.05\quad \text{for}\ \Delta T = 0.
 \end{align}
This bound constrains part of the parameter space in our model. At future lepton colliders, the sensitivity of electroweak precision observables to new physics is expected to be enhanced. Dark fermion contributions to $S$ can thus be probed if~\cite{Fan:2014vta}
\begin{align}
|\Delta S| > 0.01\quad \text{for } \Delta T = 0.
\end{align}
As we will see in Section~\ref{sec:future}, indirect searches for dark fermions in electroweak precision observables are thus competitive with Higgs couplings in certain regions of the parameter space. A detailed analysis of electroweak precision observables at future lepton colliders in the context of fermionic Higgs portals can also be found in Ref.~\cite{Fedderke:2015txa}. 

\section{Vacuum stability}\label{sec:vacuum}
\noindent New fermions with Yukawa couplings to the Higgs field will generally have an impact on the stability of the electroweak vacuum. In the symmetric phase, the Higgs potential in the SM is given by
\begin{align}
V(H) = -\frac{M_h^2}{2}\big(H^\dagger H\big) + \frac{\lambda}{2}\big(H^\dagger H\big)^2,
\end{align}
where $\lambda$ is the quartic Higgs coupling. After electroweak symmetry breaking, the triple and quartic Higgs couplings are related through $\lambda_3 = 3\lambda v$. Modifications of the triple Higgs coupling are thus directly related to the form of the Higgs potential at high energies. We consider the vacuum as stable up to a certain energy scale $\Lambda_{\rm UV}$, if the quartic coupling $\lambda(t=\log\Lambda)$ remains positive at all scales $\Lambda < \Lambda_{\rm UV}$. 

In order to study the impact of dark fermions on the vacuum stability in our model, we consider the renormalization group evolution (RGE) of the quartic Higgs coupling. At the leading order, the RGE for the relevant couplings in the SM extended by the dark fermion fields from Eq.~(\ref{eq:fields}) is given by
\begin{align}\label{eq:rge}
\frac{d\lambda}{dt} & = \frac{1}{16\pi^2}\Big(12\lambda^2 + (12y_t^2 + 8y^2 - 9g_2^2 - 3g_1^2)\lambda -12y_t^4 - 16y^4\Big),\\\nonumber
\frac{dy_t}{dt} & = \frac{y_t}{16\pi^2}\Big(\tfrac{3}{2}y_t^2 + 3y_t^2 + 2y^2 -8g_3^2 - \tfrac{9}{4}g_2^2 - \tfrac{17}{12}g_1^2\Big),\\\nonumber
\frac{dy}{dt} & = \frac{y}{16\pi^2}\Big(\tfrac{5}{2}y^2 + 3y_t^2 + 2y^2 + 2y^2 - \tfrac{9}{4}g_2^2 - \tfrac{3}{4}g_1^2\Big),\\\nonumber
\frac{dg_1}{dt} & = + \frac{15}{2}\frac{g_1^3}{16\pi^2},\quad \frac{dg_2}{dt} = - \frac{5}{2}\frac{g_2^3}{16\pi^2}, \quad \frac{dg_3}{dt} = - 7\frac{g_3^3}{16\pi^2},
\end{align}
where $g_1=g_Y$, $g_2=g$, and $g_3$ are the couplings corresponding with the $U(1)_Y$, $SU(2)_L$, and $SU(3)_C$ gauge groups, respectively, and $y_t$ is the top-quark Yukawa coupling. We have used the SM contributions from Refs.~\cite{Machacek:1983tz,Machacek:1983fi,Machacek:1984zw} and neglected the small impact of light fermions on the RGE. Similar scenarios have been discussed for instance in Refs.~\cite{Joglekar:2012vc,Altmannshofer:2013zba}.

\begin{figure}[t]
\centering
\includegraphics[height=2.2in]{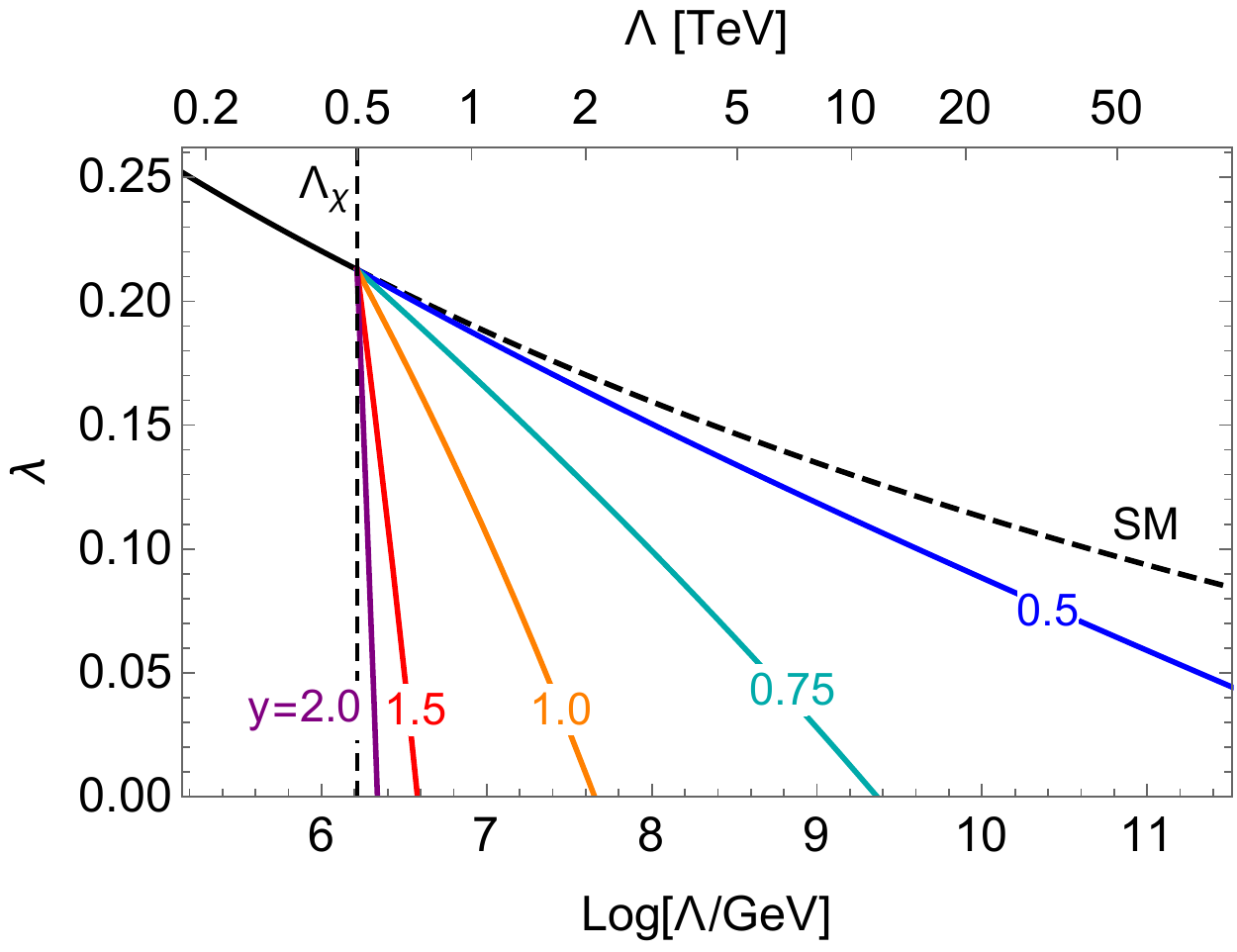} \hspace*{1cm} \includegraphics[height=2.2in]{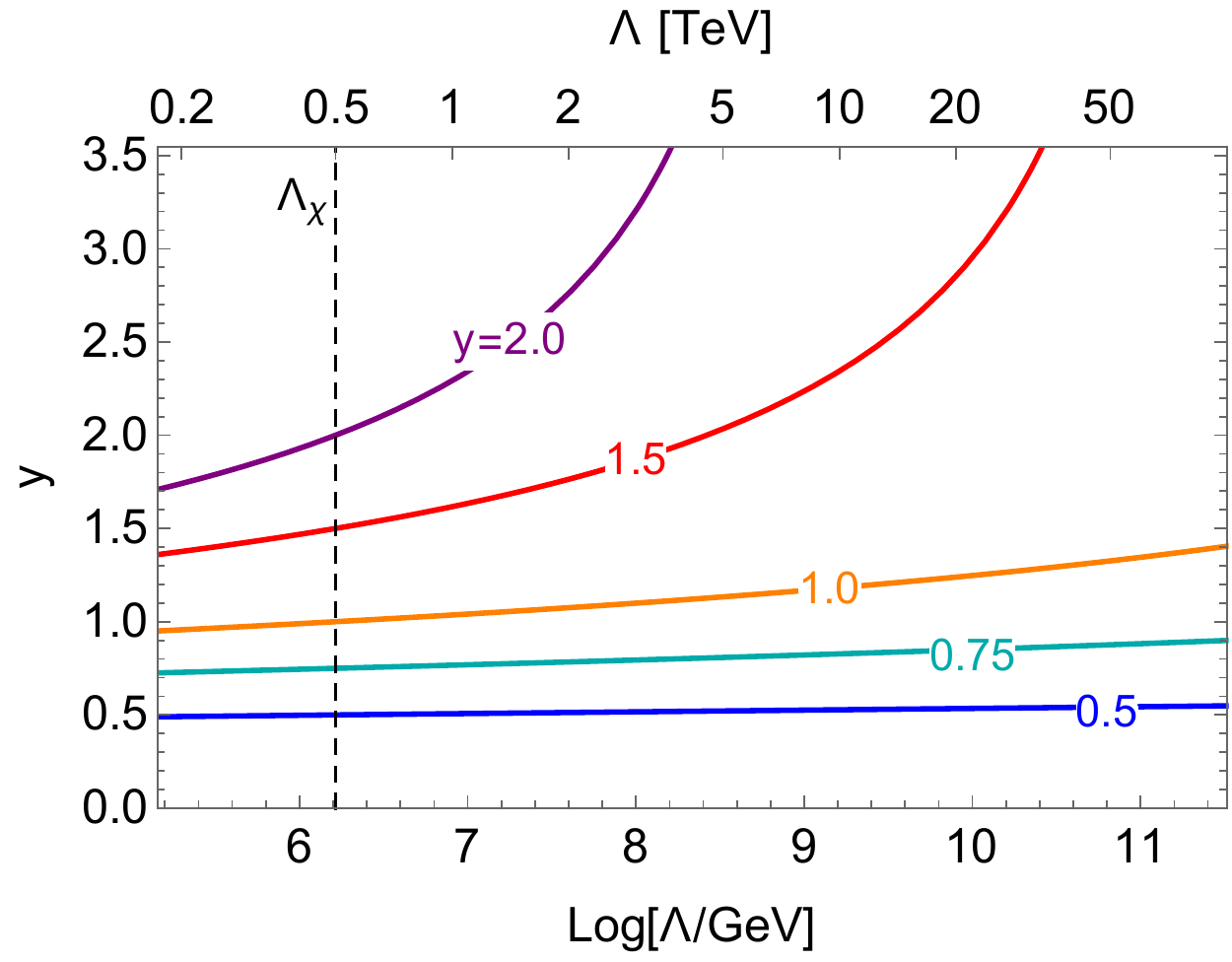}
\caption{\label{fig:vacuum}Renormalization group evolution of the quartic Higgs coupling $\lambda(t=\log\Lambda)$ (left) and the dark Yukawa coupling $y(t)$ (right) in the dark fermion model. The energy scale of the dark sector is fixed at $\Lambda_\chi = 500\,\text{GeV}$. Colored curves correspond with different values of $y(\log \Lambda_\chi)$. The dashed black curve shows the evolution of $\lambda$ in the SM.}
\end{figure}

The evolution for $\lambda$ is obtained by solving the coupled system of equations in Eq.~(\ref{eq:rge}) numerically. We use the input values for the SM couplings at the top mass scale $\Lambda = m_t$ from Ref.~\cite{Buttazzo:2013uya} and evolve the system from this scale upwards. The dark sector is assumed to set in at a single scale $\Lambda_\chi$, and threshold effects are neglected. In Fig.~\ref{fig:vacuum}, left, we show the RGE of the quartic Higgs coupling $\lambda(t)$ in our dark fermion model for various fixed values of $y(\log\Lambda_\chi)$ and a dark scale $\Lambda_\chi = 500\,\text{GeV}$. While in the SM the quartic Higgs coupling remains positive up to very high scales, in our model the quartic coupling becomes negative already below the TeV scale for strong Yukawa couplings $y\gtrsim 1.5$. The main reason for this behavior is the strong dependence of the RGE for $\lambda$ on the dark Yukawa coupling, $d\lambda/dt\sim -16y^4$. A second effect is the growth of $y$ with energy. As can be seen in Fig.~\ref{fig:vacuum}, right, starting with a large $y(\log\Lambda_\chi)\approx 2$, the dark Yukawa coupling becomes non-perturbative around $\Lambda\approx 4\,\text{TeV}$. The non-perturbative regime is thus reached at higher energies than vacuum instability.

This simple study, while far from being accurate, clearly demonstrates that a UV completion of our model is needed in order to ensure vacuum stability. It has been previously shown that various options exist to stabilize the electroweak vacuum while being in line with measurements below the TeV scale~\cite{Grojean:2004xa,Carena:2004ha}. Dark fermions with large Yukawa couplings as part of a more complete model can therefore be consistent with vacuum stability. Moreover, they might trigger a first-order electroweak phase transition that facilitates electroweak baryogenesis~\cite{Carena:2004ha,Davoudiasl:2012tu,Fairbairn:2013xaa,Chao:2015uoa}. Since potential effects of vacuum stabilization on LHC observables strongly depend on the specific UV completion of our model, we do not consider them in our analysis. However, the reader should bear in mind that such effects could naturally exist and lead to a modified or enriched phenomenology of our scenario.

\section{Dark fermions at the LHC and future lepton colliders}\label{sec:future}
\noindent In this section, we explore the sensitivity of the LHC and future lepton colliders to virtual effects of dark fermions in Higgs observables. Dark fermion effects in Higgs-$Z$ boson associated production at future lepton colliders have also been studied in Refs.~\cite{Fedderke:2015txa,Xiang:2017yfs}. Establishing indirect evidence of dark fermions will crucially depend on the achievable precision in Higgs coupling measurements. In what follows, we will make the conservative assumption that theory uncertainties on the SM predictions will remain as they are today. \\

Let us first consider Higgs-$Z$ boson interactions. At the LHC, these can be tested in Higgs decays, weak boson fusion and associated Higgs-$Z$ production. A global analysis of Higgs measurements with data from run I leads to a bound on $F_Z^{0,pp}$ (cf. Eq.~(\ref{eq:ref-hvv})) at $68\%$ C.L.~\cite{Corbett:2015ksa}\footnote{A recent measurement of $h\to ZZ^{\ast}\to 4\ell$ with run-II data leads to a similar bound~\cite{atlas-zzh-lhc}.}
\begin{align}
\text{LHC\ run\ I:}\quad -0.08 \lesssim \text{Re}[F_Z^{0,pp}] \lesssim 0.17.
\end{align}
From Fig.~\ref{fig:f0-parameters}, left, we deduce that current measurements of $hZZ$ couplings already exclude dark fermion scenarios with large Yukawa couplings $y\gtrsim 3$ and a light neutral state $\chi_1^0$ with $|m_1|\gtrsim M_h/2$. For the HL-LHC with $3\,\text{ab}^{-1}$ luminosity, the CMS and ATLAS collaborations have predicted the sensitivity to anomalous Higgs couplings from a global analysis of Higgs production and decay channels~\cite{CMS:2013xfa,atlas-hl-lhc}. Translated to our scenario, the HL-LHC will be sensitive to dark fermions in $hZZ$ interactions for
\begin{align}
\text{HL-LHC}:\quad |\text{Re}[F_Z^{0,pp}]| \gtrsim 0.033.
\end{align}
The sensitivity to $F_Z^0(M_h^2,k_1^2,M_Z^2)$ in associated Higgs-$Z$ boson production alone is slightly lower, since the form factor is probed in a different kinematic region.

\begin{figure}[t]
\centering
\includegraphics[height=2.9in]{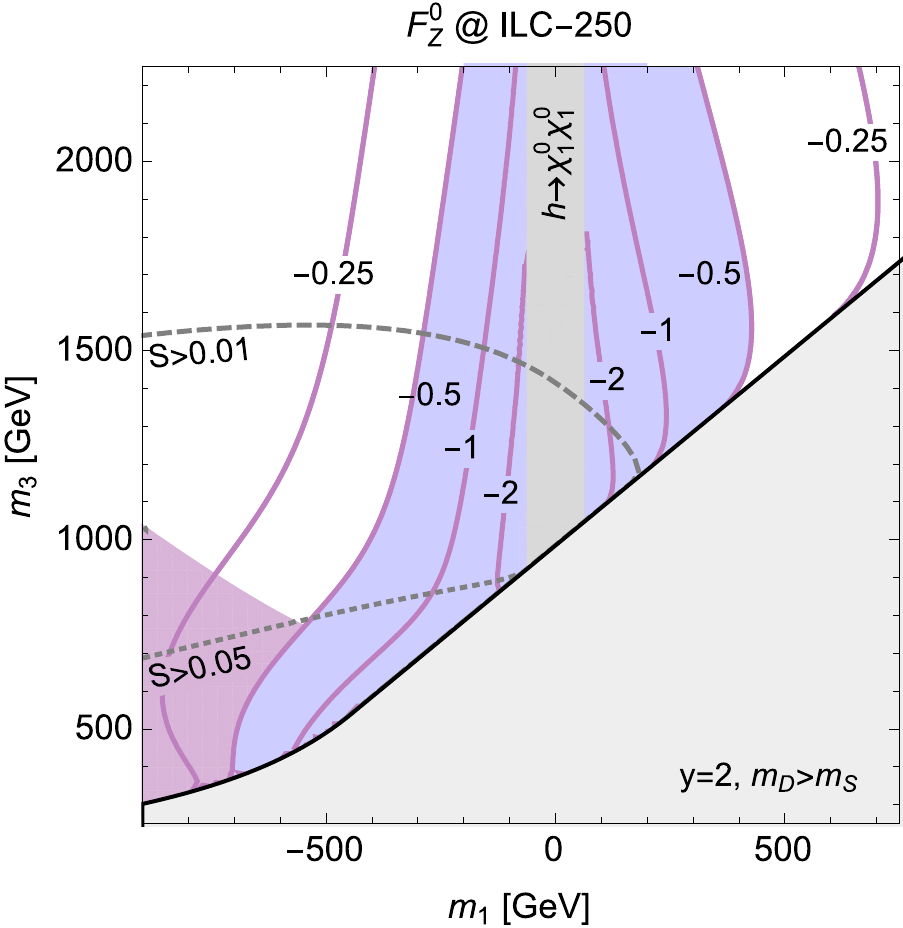} \hspace*{0.7cm} \includegraphics[height=2.9in]{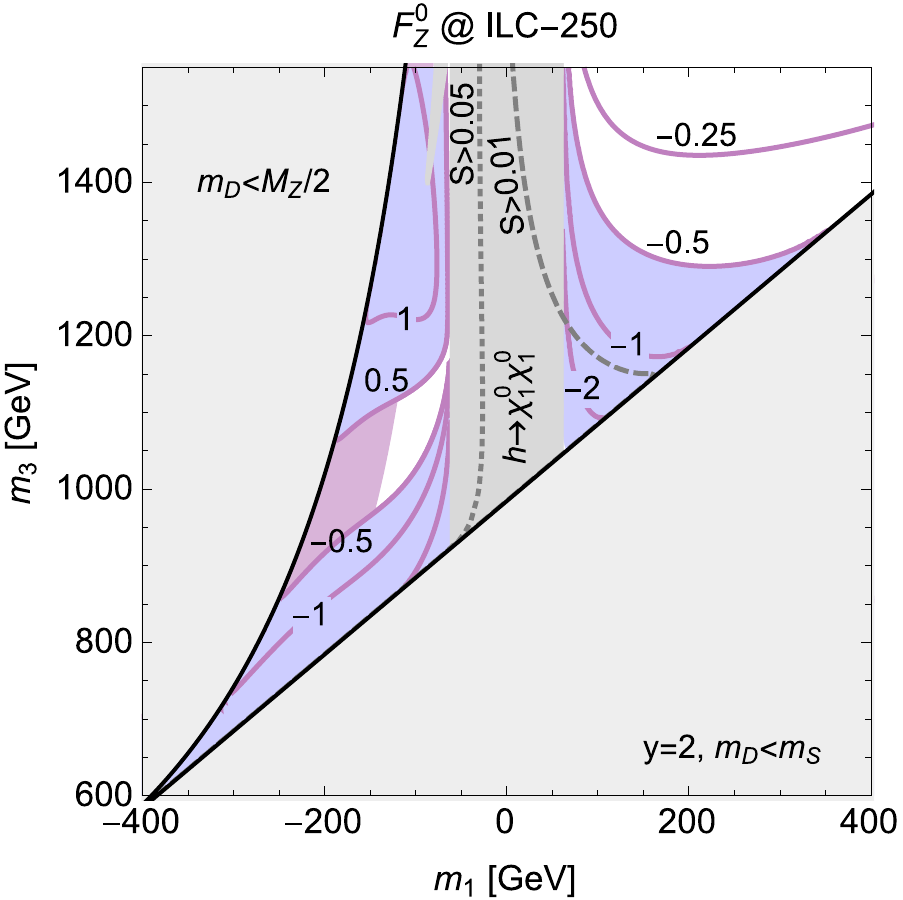}
\caption{\label{fig:f0z-mdgms}Form factor $\text{Re}[F_Z^{0,ee}]$ as a function of $m_1$ and $m_3$ for $y=2$ and $m_D > m_S$ (left) or $m_D < m_S$ (right). Purple curves show constant values of $\text{Re}[F_Z^{0,ee}]$ in percent; in purple areas, $\chi_2^0$ is the lightest state. Blue areas can be probed at future lepton colliders. Dark gray bands are excluded by LHC bounds on the Higgs width. Regions below/left of the dotted (dashed) gray lines are in tension with the $S$ parameter at LEP (can be probed at a FLC).}
\end{figure}
At future lepton colliders, the Higgs coupling to $Z$ bosons can be measured very precisely in associated production via $e^+e^-\to Zh$. Recent studies predict an uncertainty of less than half a percent for all considered machine designs~\cite{Durieux:2017rsg,Lafaye:2017kgf}. Virtual dark fermions can thus be probed for
\begin{align}
e^+e^-\to Zh:\quad |\text{Re}[F_Z^{0,ee}]| \gtrsim 0.005.
\end{align}
To illustrate the reach of a future lepton collider in dark fermion searches, we investigate the form factor in terms of the masses $m_1$ and $m_3$. In Fig.~\ref{fig:f0z-mdgms}, we show $F_Z^{0,ee}$ for a fixed dark Yukawa coupling $y=2$ in the two parameter regions $m_D > m_S$ (left) and $m_D < m_S$ (right). Constant values of $F_Z^{0,ee}$ in percent are shown as purple curves. In purple regions, $\chi_2^0$ is the lightest dark fermion state. In regions where $\chi_1^0$ is the lightest state, the heaviest state, $\chi_3^0$, decouples as a consequence of the mass splitting due to the large Yukawa coupling. Light gray regions are excluded either because $m_3-m_1 < 4 v$, which is unphysical, or because $m_D < M_Z/2$, which is strongly constrained by measurements of the $Z$ boson width. Gray regions have been excluded by bounds on invisible Higgs decays $h\to \chi_1^0\chi_1^0$ from LHC run I. Electroweak precision measurements at LEP challenge the region below/left to the dotted gray line, where contributions to the $S$ parameter are sizable. As the mass hierarchy $m \gg v$ entering the definition of $S$ is not fulfilled in all regions of parameter space, we cannot claim a strict exclusion, but consider these regions in tension with LEP measurements.

A FLC with $\sqrt{s} = 250\,\text{GeV}$ can explore the blue parameter regions in associated production $e^+e^-\to Zh$. In this process, Higgs-$Z$ couplings will be efficient probes of dark fermions if either of the lightest states, $\chi_1^0$ or $\chi_2^0$, lies in the mass range
\begin{align}\label{eq:hzz-reach}
m_D > m_S: & \quad -550\,\text{GeV}\lesssim m_1 \lesssim 400\,\text{GeV} \quad \text{or} \quad m_2 \lesssim 550\,\text{GeV},\\\nonumber
m_D < m_S: & \quad -200\,\text{GeV} \lesssim m_1 \lesssim 350\,\text{GeV} \quad \text{or} \quad m_2 \lesssim 200\,\text{GeV}.
\end{align}
With improved experimental sensitivity and/or reduced theory uncertainties, the mass reach of anomalous Higgs couplings can be extended to adjacent regions. The sensitivity of the $S$ parameter is expected to reach up to the dashed gray line. Electroweak precision measurements will thus cover the entire parameter region of scenario 2, where $\chi_2^0$ is the lightest state. Higgs observables complement and surpass electroweak observables, especially in scenario 1 with $\chi_1^0$ as the lightest state.\\

Turning to triple Higgs interactions, the currently strongest direct bound has been obtained by the CMS collaboration from an analysis of Higgs pair production with the subsequent decay $hh\to (b\bar b) (\gamma\gamma)$~\cite{cms-di-higgs}. It is based on $36\,\text{fb}^{-1}$ of LHC data collected at $\sqrt{s} = 13\,\text{TeV}$. Applied to our model, the result translates into a bound on the form factor $\overline{F}_h$ (cf. Eq.~(\ref{eq:ref-hhh})) at $95\%$~C.L.,
\begin{align}
pp\to hh\to (b\bar b)(\gamma\gamma):\quad -10 \lesssim \text{Re}[\overline{F}_h] \lesssim 14.
\end{align}
The asymmetric sensitivity to $F_h$ is due to the negative interference of the signal and background amplitudes $gg\to h^\ast \to hh$ and $gg\to hh$ in the SM, where the latter does not involve the triple Higgs coupling. A negative form factor $F_h$ thus causes a positive correction to this interference term, which enhances the sensitivity to the Higgs pair signal. Comparing with Fig.~\ref{fig:f0h-parameters}, left, it is apparent that with the current precision Higgs pair production is not sensitive to dark fermions with perturbative couplings yet. For the HL-LHC, the ATLAS collaboration has predicted the sensitivity to $\lambda_3$ in Higgs pair production in the decay channels $hh\to (b\bar b)(\tau^+\tau^-)$~\cite{atlas-di-higgs-hl-lhc-tata} and $hh\to (b\bar b)(\gamma\gamma)$~\cite{atlas-di-higgs-hl-lhc-gaga}. The results translate into the following ranges that can be probed at the HL-LHC,
\begin{align}
pp\to hh\to (b\bar b)(\tau^+\tau^-):\quad & \text{Re}[\overline{F}_h] \lesssim -5 \quad \text{or} \quad \text{Re}[\overline{F}_h] \gtrsim 11,\\\nonumber
pp\to hh\to (b\bar b)(\gamma\gamma):\quad & \text{Re}[\overline{F}_h] \lesssim -2 \quad \text{or} \quad \text{Re}[\overline{F}_h] \gtrsim 7.
\end{align}
As can be observed from Fig.~\ref{fig:f0h-parameters}, left, Higgs pair production at the HL-LHC should probe dark fermions with couplings $y\gtrsim 2$.

\begin{figure}[t]
\centering
\includegraphics[height=2.9in]{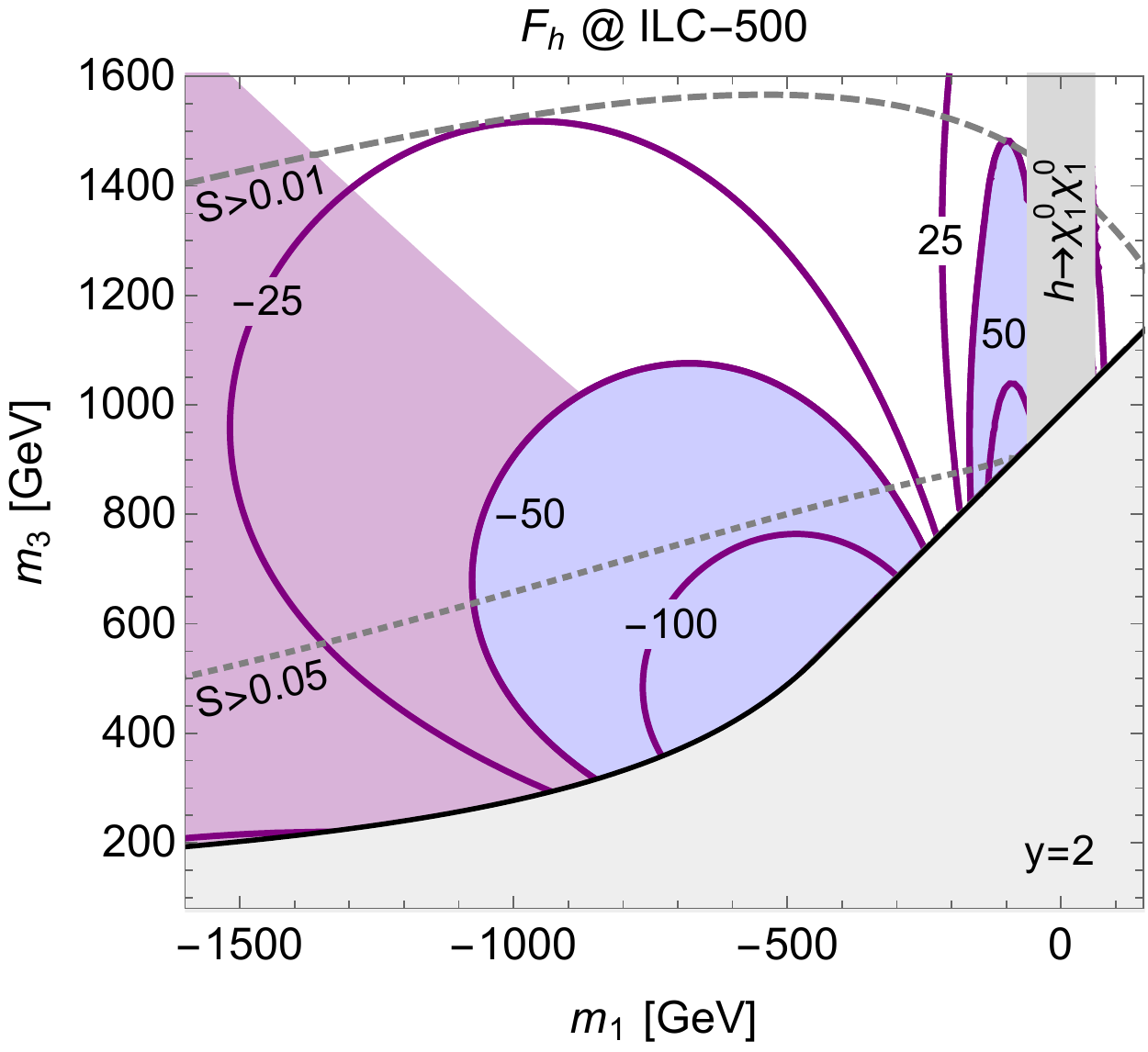}
\caption{\label{fig:f0h-mdgms}Form factor $\text{Re}[\overline{F}_h]$ as a function of $m_1$ and $m_3$ for fixed $y=2$. Purple curves show constant values of $\overline{F}_h$ in percent; in purple areas, $\chi_2^0$ is the lightest state. Blue areas can be probed at a future lepton collider with $\sqrt{s} = 500\,\text{GeV}$. Dark gray areas are excluded by LHC bounds on the Higgs invisible width. Regions below the dotted (dashed) gray lines are in tension with the $S$ parameter at LEP (can be probed at a FLC).}
\end{figure}
At a future lepton collider with $\sqrt{s} > 2M_h+M_Z$, the triple Higgs coupling can be probed through the process $e^+ e^-\to Z^\ast \to Zh^\ast \to Zhh$~\cite{Djouadi:1999gv}. The sensitivity will crucially depend on how well the signal with triple Higgs interactions can be discriminated from irreducible SM background. At the ILC with $\sqrt{s} = 500\,\text{GeV}$, current estimates with $h\to b\bar b$ decays lie in the range $\Delta \lambda_3/\lambda_3\approx 30-50\%$~\cite{Asner:2013psa}. For our predictions, we will make the rather conservative assumption of probing
\begin{align}
e^+e^-\to Zhh:\quad |\text{Re}[\overline{F}_h]| \gtrsim 0.5. 
\end{align}
 A similar sensitivity is expected from a future 100-TeV proton-proton collider probing anomalous triple Higgs couplings in Higgs pair production~\cite{Contino:2016spe}. In Fig.~\ref{fig:f0h-mdgms}, we illustrate the sensitivity of a future lepton collider with $\sqrt{s}=500\,\text{GeV}$ to virtual dark fermions in triple Higgs couplings. Displayed is the form factor $\overline{F}_h$ in terms of $m_1$ and $m_3$ for fixed $y=2$. Notice that $F_h$ is symmetric under $m_D\leftrightarrow m_S$, so that the entire parameter space can be displayed in one panel. Bounds from the $S$ parameter are shown for $m_D > m_S$; for $m_D < m_S$, they can be obtained from Fig.~\ref{fig:f0z-mdgms}, right. Comparing Figs.~\ref{fig:f0z-mdgms} and~\ref{fig:f0h-mdgms}, we see that in the blue regions accessible at a FLC, effects of dark fermions in triple Higgs couplings are typically two orders of magnitude larger than in Higgs-gauge couplings. This enhancement is compensated by the different observation prospects, so that the sensitivity of both anomalous couplings is comparable in this region. In parameter regions where $\chi_1^0$ or $\chi_2^0$ are the lightest states, respectively, dark fermions in triple Higgs interactions can be probed in the mass range
 \begin{align}
-900\,\text{GeV} \lesssim m_1 \lesssim -250\,\text{GeV}\ \cup\ -180\,\text{GeV} \lesssim m_1 \lesssim 0\,\text{GeV} \quad \text{or} \quad m_2 \lesssim 900\,\text{GeV}.
\end{align}
Compared with Higgs-$Z$ couplings, see Eq.~(\ref{eq:hzz-reach}), triple Higgs couplings can thus probe scenarios with an overall heavier spectrum, where the lightest new states lie close to the TeV scale. Moreover, in triple Higgs interactions, the sensitivity reaches further into the region where $\chi_2^0$ is the lightest state. This region, in turn, can also be tested in electroweak precision measurements. From Fig.~\ref{fig:f0h-mdgms}, it is apparent that the entire parameter space accessible in triple Higgs couplings can also be probed by the $S$ parameter at a FLC.

Complementary to direct measurements, triple Higgs interactions could also be probed indirectly through higher-order corrections to single Higgs production processes~\cite{Degrassi:2016wml,Bizon:2016wgr,McCullough:2013rea}. While the sensitivity to $\lambda_3$ can be comparable with direct observables, predictions of the indirect observables can be modified by virtual effects occurring one loop order before the triple Higgs modification. For instance, at a future lepton collider running below the threshold of resonant Higgs pair production, anomalous triple Higgs interactions can be tested in $e^+e^-\to hZ$ through electroweak corrections~\cite{McCullough:2013rea}. In our scenario, anomalous $hhh$ couplings at NLO are competing with anomalous $hZZ$ at LO in this process. An ana\-ly\-sis of dark fermions in indirect triple Higgs observables therefore requires dedicated calculations beyond the scope of this work.

\section{Comparison with resonant dark fermion production}\label{sec:resonant}
\noindent In order to establish the complementarity of indirect and direct collider searches for dark fermions, we compare our results on anomalous Higgs couplings with resonant dark fermion production at the LHC and future lepton colliders. At the LHC, dark fermions can be pair-produced through electroweak interactions via\begin{align}
pp\to V^\ast \to \chi_i\chi_j,
\end{align}
and subsequently decay into lighter fermions and SM bosons. The dominant production and decay channels are determined by the gauge couplings and the masses of the dark fermions. Due to the large dark Yukawa coupling, the mass difference between $\chi_1^0$ and $\chi_3^0$ is typically sizable, so that $\chi_3^0$ is often too heavy to be pair-produced at observable rates. The phenomenology of resonant dark fermion production is thus dominated by the lighter states $\chi_1^0$ and $\chi_2^0$.\footnote{In parameter regions with large negative $m_1$, the mass hierarchy can be inverted, so that $\chi_3^0$ and $\chi_2^0$ can be produced resonantly at the LHC.} Effects of virtual dark fermions in Higgs couplings are large if the coupling $h\chi_1^0\chi_1^0\sim \sin(2\theta)$ is sizable (see Eq.~(\ref{eq:higgs-gauge})). Maximal Higgs effects are thus expected for
\begin{align}\label{eq:max-higgs}
\sin(2\theta)=1,\quad \cos(2\theta) = 0,\quad \sin\theta = 1/\sqrt{2} = \cos\theta.
\end{align}
In this limit, the off-diagonal Higgs coupling $h\chi_1^0\chi_3^0\sim \cos(2\theta)$ is absent, and gauge couplings to $\chi_1^0$ and $\chi_3^0$ are of the same strength. We adopt the maximal-Higgs limit for our analysis of resonant fermion production. As we will see, it leads to a characteristic pattern of signatures, from which we will determine the parameter regions where Higgs couplings perform better than direct searches.

In scenario 1, where $|m_1| < m_2\lesssim m_c < m_3$, the dominant production channels in the maximal-Higgs limit from Eq.~(\ref{eq:max-higgs}) are
\begin{align}
pp\to Z^\ast \to \chi_1^0\chi_2^0,\quad Z^\ast \to \chi^+\chi^-,\quad W^\ast \to \chi^\pm\chi_1^0,\quad W^\ast \to \chi^\pm \chi_2^0.
\end{align}
The so-produced mediator states decay via
\begin{align}
\chi_2^0\to Z\chi_1^0,\quad \chi^\pm \to W^\pm \chi_1^0.
\end{align}
The main signals at the LHC are thus made of gauge bosons and missing energy, $E_T^{\text{miss}}$,
\begin{align}\label{eq:lhc-signals}
Z + E_T^{\text{miss}}: & \qquad pp\to Z^\ast \to \chi_2^0\chi_1^0\to (Z\chi_1^0)\chi_1^0,\\\nonumber
W + E_T^{\text{miss}}: & \qquad pp\to W^\ast \to \chi^\pm \chi_1^0 \to (W^\pm \chi_1^0)\chi_1^0,\\\nonumber
WW + E_T^{\text{miss}}: & \qquad pp\to Z^\ast \to \chi^+ \chi^- \to (W^+\chi_1^0)(W^-\chi_1^0),\\\nonumber
WZ + E_T^{\text{miss}}: & \qquad pp\to W^\ast \to \chi^\pm \chi_2^0 \to (W^\pm \chi_1^0)(Z\chi_1^0).
\end{align}
For $m_D - |m_1| > M_V$, the final gauge bosons are produced resonantly and can be detected through their decays into leptons and jets~\cite{Calibbi:2015nha}. Current searches for signatures with two bosons and missing energy in the context of supersymmetry are sensitive to masses of the lightest state up to $100-200\,\text{GeV}$, if the next-to-lightest states lie below about $500\,\text{GeV}$~\cite{ATLAS:2017uun,CMS:2017sqn}. If the mass splitting drops below the threshold of resonant gauge boson production, decay products are soft and more difficult to observe~\cite{Gori:2013ala,Schwaller:2013baa,CMS:2017fij}. For the $W+E_T^{\text{miss}}$ signature, the small signal rate is overwhelmed by SM background and probably not observable at the LHC. The situation is better for $Z+E_T^{\text{miss}}$, where the reconstruction of a lepton pair might facilitate the observation of a signal~\cite{Enberg:2007rp}. From mono-jet searches, we do not expect much additional information, due to the small production rates of invisible final states.

In scenario 2 with $m_2 \lesssim m_c < |m_1| < m_3$, the production of dark fermions proceeds as in scenario 1. The dominant decay channels are
\begin{align}
\chi_1^0\to Z\chi_2^0,\quad \chi_1^0\to W^\pm \chi^\mp,\quad \chi^\pm \to W^\ast \chi_2^0\to (f\bar{f}')\chi_2^0.
\end{align}
The main LHC signatures in scenario 2 are as in Eq.~(\ref{eq:lhc-signals}) with $\chi_1^0\leftrightarrow \chi_2^0$. However, due to the small mass splitting between the doublet states, $m_c - m_2 \lesssim 1\,\text{GeV}$, the decay of $\chi^\pm$ always proceeds through an off-shell $W$ boson. The decay products from $WW + E_T^{\text{miss}}$ and $WZ + E_T^{\text{miss}}$ final states are thus too soft to be observed even in dedicated searches for soft leptons, which require a minimum transverse momentum of $5\,\text{GeV}$~\cite{CMS:2017fij}. In the future, disappearing charged tracks might offer a possibility to search for such scenarios in regions with a sufficiently small mass splitting between $\chi_1^0$ and $\chi^\pm$~\cite{Mahbubani:2017gjh}. $Z+E_T^{\text{miss}}$ might be an alternative way to observation. In summary, scenario 2 is more difficult to test through resonant production than scenario 1 in most of the parameter space.

While the sensitivity of current resonant searches is restricted to parameter regions where $\chi_1^0$, $\chi_2^0$, and $\chi^\pm$ are all relatively light, anomalous Higgs-gauge couplings can probe regions with heavier doublet states $\chi_2^0,\chi^\pm$ (cf. Fig.~\ref{fig:f0-parameters}, left). As of today, both approaches are thus complementary in their sensitivity to dark fermions. At the HL-LHC, the reach of resonant searches is expected to be comparable with what is observed today, as it is basically determined by the available collider energy. Anomalous Higgs-gauge couplings start testing scenarios with smaller Yukawa couplings (i.e., smaller mass splittings) that can be probed by resonance searches as well (see Fig.~\ref{fig:f0-parameters}, left). Triple Higgs couplings will become available as additional probes of the region with large $\chi_1^0-\chi_2^0$ splitting (see Fig.~\ref{fig:f0h-parameters}, left). At a future $100$-TeV proton-proton collider, searches for charged leptons and missing energy can eventually probe dark fermions up to the TeV scale~(a summary can be found in Ref.~\cite{Freitas:2015hsa}).

Complementary to electroweak production, dark fermions might be explored through off-shell Higgs production and subsequent decay into dark fermion pairs, $gg\to h^\ast \to \chi_1^0\chi_1^0$. At hadron colliders, the do\-mi\-nant channels in scenario 1 are mono-jet production, weak boson fusion, and Higgs-associated top-antitop production. Due to the strong suppression by the small Higgs width, however, such processes are rare and difficult to observe. The sensitivity of the (HL-)LHC and a future $100$-TeV collider has been analyzed in similar Higgs-portal scenarios and predicted to be weak~\cite{Craig:2014lda}. In scenario 2, the lightest state $\chi_2^0$ does not couple to the Higgs boson, so that off-shell Higgs observables are not an option to probe this case.

At lepton colliders, the dominant process to produce resonant dark fermions is
\begin{align}
e^+e^-\to Z^\ast \to \chi^+\chi^-.
\end{align}
The LEP collaborations have searched for pairs of heavy charged fermions $F^\pm$, produced via $e^+e^-\to Z^\ast \to F^+ F^-$ and decaying through $F^\pm \to W^\pm F^0$, where $F^0$ is a stable neutral fermion. In scenario 1 of our model, the null results of searches for leptons and missing energy constrain the mass of charged dark fermions decaying via $\chi^\pm\to W^\pm \chi_1^0$ to~\cite{Abdallah:2003xe}
\begin{align}
m_c \approx m_2 \approx m_D \gtrsim 100\,\text{GeV},\quad \text{if  }m_c - |m_1| \gtrsim 5\,\text{GeV}. 
\end{align}
In the parameter regions where anomalous Higgs couplings are sizable, this bound is mostly irrelevant, apart from a narrow excluded region of $F_Z^{0,pp}$ for $m_D < m_S$ (see Fig.~\ref{fig:f0z-mdgms}, right). A future lepton collider with higher energy $\sqrt{s} > 200\,\text{GeV}$ could extend the reach to charged dark fermions with masses
\begin{align}
m_c \lesssim \sqrt{s}/2.
\end{align}
Provided that the decay products can be detected, a FLC with $\sqrt{s}=250\,\text{GeV}$ will probe the region where $F_Z^{0,ee} > 0$ and $|m_1| < m_2$ (see Fig.~\ref{fig:f0z-mdgms}, right). Similarly, a FLC with $\sqrt{s}=500\,\text{GeV}$ should cover most of the region where $\overline{F}_h > 0$ (see Fig.~\ref{fig:f0h-mdgms}). In scenario 2, where $m_c - m_2 < 1\,\text{GeV}$, the decay products from $\chi^\pm \to W^\pm \chi_2^0$ are too soft to be detected, but the mass splitting is typically also too large for a displaced vertex or charged track of $\chi^\pm$. In this case, invisibly decaying charged fermions lead to mono-photon signatures. These have been analyzed at LEP in the context of the MSSM, excluding charginos with masses below $75\,\text{GeV}$~\cite{Abdallah:2003xe}. Dedicated searches for mono-photon signatures at a FLC could thus help exploring regions with $m_c - m_2 < 1\,\text{GeV}$.

In conclusion, in the regime of large Yukawa couplings dark fermions appear to hide from direct observation. The sensitivity of resonant dark fermion production is limited by the small production rates especially for heavier mediator states and/or detection inefficiencies in the case of compressed spectra. Virtual effects, in turn, are typically less dependent on the mediators, but rather determined by the lightest states of the spectrum. Indirect searches for such effects can indeed probe scenarios that are inaccessible in direct searches.

\section{Conclusions}\label{sec:conclusions}
\noindent Let us summarize the status and prospects of indirect collider searches for dark fermions as follows. In scenarios with large Yukawa couplings, Higgs-gauge and triple Higgs interactions are sensitive to virtual contributions of dark fermions. We have analyzed in detail the singlet-doublet model of Majorana fermions, which has the same particle content as the bino-higgsino scenario in the MSSM, but features large fermion mixing. The main effects on Higgs observables are a correction of the Higgs couplings to $W$ and $Z$ bosons, which is typically negative, and a modification of the triple Higgs vertex, which can have either sign.

Anomalous Higgs couplings are actively being investigated in Higgs production and decay at the LHC. Numerically, virtual corrections of dark fermions in triple Higgs couplings are one to two orders of magnitude larger than in Higgs-gauge couplings. The latter, in turn, can be measured much more precisely. This over-compensates the suppression in magnitude, rendering Higgs-gauge interactions a more sensitive probe of dark fermions at hadron colliders. A global analysis of Higgs observables with run-I data has already excluded dark fermions with strong Yukawa couplings $y\gtrsim 3$ through modified Higgs-$Z$ boson interactions (Fig.~\ref{fig:f0-parameters}, left). At the HL-LHC with $3\,\text{ab}^{-1}$ of data, Higgs-gauge couplings are expected to probe dark fermions for $y\gtrsim 2$. Triple Higgs couplings will be somewhat less sensitive, covering scenarios with $y\gtrsim 2.5$ (Fig.~\ref{fig:f0h-parameters}, left).

For dark fermions with smaller Yukawa couplings $y\gtrsim 1$, a future lepton collider will be an excellent test ground. Precise measurements of the Higgs-gauge coupling in Higgs-$Z$ associated production will allow us to probe dark fermions up to the TeV scale (Fig.~\ref{fig:f0z-mdgms}). This can be achieved with basically all currently discussed machine designs. At a lepton collider running at or above $\sqrt{s} = 500\,\text{GeV}$, triple Higgs couplings measured through $Z$-associated Higgs pair production can test dark fermions in a largely complementary range of parameter space (Fig.~\ref{fig:f0h-mdgms}). At high-energy lepton colliders, the combination of Higgs-gauge and triple Higgs interactions is thus a powerful strategy to explore possible scenarios of dark fermions. Higgs observables are complementary with (and in some regions superior to) electroweak precision observables not involving the Higgs boson. 

Compared with direct collider searches for resonant dark fermions, anomalous Higgs couplings can cover regions where mediator states are too heavy to be produced at observable rates, both at hadron and even more so at lepton colliders. Importantly, indirect observables are also sensitive to scenarios with compressed spectra, where decay products of dark fermions are difficult to detect. This example teaches us to pursue a two-fold search strategy for dark sectors at colliders: through resonant production and through virtual effects. It might well be that dark particles hidden from direct searches will first show up as quantum corrections in precision observables.

\section{Acknowledgments}
\noindent We thank Michael Kr\"amer, Luminita Mihaila, Tilman Plehn, Pedro Schwaller, Jamie Tattersall, and Jos\'e Zurita for helpful discussions. AV would like to thank the Institute for Theoretical Physics in Heidelberg for its warm hospitality. SW wants to thank the Mainz Institute for Theoretical Physics (MITP) for its hospitality and support, where part of this work was completed. We furthermore acknowledge support by the DFG Research Unit \emph{New Physics at the LHC} (FOR2239). SW acknowledges funding by the Carl Zeiss Foundation through a \emph{Junior-Stiftungsprofessur}.

\appendix

\section{One-loop vertex corrections from heavy vector-like fermions}\label{app:hvv}
\noindent In this appendix, we give analytic results of virtual corrections to Higgs-gauge and triple Higgs interactions from new fermions. While the explicit expressions correspond to our model of dark fermions, the loop functions apply more generally to massive fermions with Yukawa couplings and vector-like weak interactions.\\

\begin{table}
\centering
\begin{tabular}{ccc}\label{tab:couplings}
\begin{tabular}{c|cccc}
$g^{ij}_Z$ & $\chi_1^0$ & $\chi_2^0$  & $\chi_3^0$ & $\chi^+$\\
\hline
$\chi_1^0$ & $0$ & $+\frac{e}{2s_Wc_W}\cos\theta$ & $0$ & $0$\\
$\chi_2^0$ & $-\frac{e}{2s_Wc_W}\cos\theta$ & $ 0$ & $+\frac{e}{2s_Wc_W}\sin\theta$ & $0$\\
$\chi_3^0$ & $0$ & $-\frac{e}{2s_Wc_W}\sin\theta$ & $0$ & $0$\\
$\chi^-$ & $0$ & $0$ & $0$ & $-i\frac{e}{s_Wc_W}(\frac{1}{2}-s_W^2)$\\
\end{tabular}\hspace*{0.4cm}
&
\raisebox{0.25cm}{\begin{tabular}{c|c}
$g^{ic}_{W}$ & $\chi^+$\\
\hline
$\chi_1^0$ & $+i\frac{e}{2s_W}\cos\theta$\\
$\chi_2^0$ & $+\frac{e}{2s_W}$\\
$\chi_3^0$ & $-i\frac{e}{2s_W}\sin\theta$\\
\end{tabular}}\hspace*{0.4cm}
&
\raisebox{0.25cm}{\begin{tabular}{c|c}
$g^{cj}_{W}$ & $\chi^-$\\
\hline
$\chi_1^0$ & $+i\frac{e}{2s_W}\cos\theta$\\
$\chi_2^0$ & $-\frac{e}{2s_W}$\\
$\chi_3^0$ & $-i\frac{e}{2s_W}\sin\theta$\\
\end{tabular}}
\end{tabular}
\caption{Coupling strength of vector-like Weyl fermions to weak gauge bosons in our dark fermion model. Here $g_V^{ij}$ denotes the coupling of boson $V=W,Z$ to an incoming fermion in mass eigenstate $i=\{1,2,3,c\}$ and an outgoing fermion in mass eigenstate $j=\{1,2,3,c\}$.}
\end{table}
In our model, the Yukawa couplings of dark fermions to the Higgs boson read
\begin{align}
g^{11}_h & = iy \sin(2\theta),\quad g^{13}_h = iy \cos(2\theta) = g^{31}_h,\quad g^{33}_h = -iy \sin(2\theta),\quad g_h^{2i} = 0 = g_h^{ci}.
\end{align}
The gauge couplings of dark fermions to $Z$ and $W$ bosons are given in Table~\ref{tab:couplings}. The amplitudes of the one-loop diagrams have been calculated with a private computer code based on the programs {\tt FeynArts}~\cite{Hahn:2000kx}, {\tt FeynCalc}~\cite{Mertig:1990an}, and {\tt LoopTools}~\cite{Hahn:1998yk}, and cross-checked using the computer tool {\tt SARAH}~\cite{Staub:2013tta}. The loop functions for the $hVV$ vertex corrections are given by
\begin{align}
L_V^0(p^2, k_1^2, k_2^2, m_i,m_j,m_D) & = (m_i + m_j)B_0[p^2, m_i^2, m_j^2] + (m_i - m_D)B_0[k_1^2, m_i^2, m_D^2]\\\nonumber
& + (m_j - m_D)B_0[k_2^2, m_j^2, m_D^2] - 4(m_i + m_j)C_{00}[p^2,k_2^2, k_1^2, m_i^2, m_j^2, m_D^2]\\\nonumber
& \hspace*{-4cm} + \big\{m_D p^2  - m_j k_1^2 - m_i k_2^2 + (m_i+m_j)(m_D^2 - m_D (m_i+m_j) + m_im_j) \big\} C_0[p^2, k_1^2, k_2^2, m_j^2, m_i^2, m_D^2],\\
L_V^1(p^2, k_1^2, k_2^2, m_i,m_j,m_D)/M_V^2 & = 2(m_D - m_i)C_2[p^2, k_2^2, k_1^2, m_i^2, m_j^2, m_D^2]\\\nonumber
& - 2(3m_i + m_j)C_1[p^2,k_2^2, k_1^2, m_i^2, m_j^2, m_D^2] - 2m_i C_0[p^2,k_1^2, k_2^2, m_j^2, m_i^2, m_D^2]\\\nonumber
 & - 4(m_i+m_j)\big\{C_{11}[p^2, k_2^2, k_1^2, m_i^2, m_j^2, m_D^2] + C_{12}[p^2, k_2^2, k_1^2, m_i^2, m_j^2, m_D^2]\big\},\\
 \widetilde{L}_V^1(p^2, k_1^2, k_2^2, m_i,m_j,m_D)/M_V^2 & = 2(m_i + 3m_j)C_2[p^2, k_2^2, k_1^2, m_i^2, m_j^2, m_D^2]\\\nonumber
& - 2(m_D - m_j)C_1[p^2,k_2^2, k_1^2, m_i^2, m_j^2, m_D^2] + 2m_j C_0[p^2,k_1^2, k_2^2, m_j^2, m_i^2, m_D^2]\\\nonumber
 & + 4(m_i+m_j)\big\{C_{22}[p^2, k_2^2, k_1^2, m_i^2, m_j^2, m_D^2] + C_{12}[p^2, k_2^2, k_1^2, m_i^2, m_j^2, m_D^2]\big\}.
\end{align} 
The loop function for the triple Higgs vertex correction is given by
\begin{align}
L_h(p^2, k_1^2, k_2^2, m_i,m_j) & = 2m_iB_0[p^2, m_i^2, m_i^2] + (m_i + m_j)\big\{B_0[k_1^2, m_i^2, m_j^2] + B_0[k_2^2, m_i^2, m_j^2]\big\}\\\nonumber
& - \big\{m_j p^2 + m_i(k_1^2+k_2^2) - 2m_i (m_i + m_j)^2\big\}C_0[p^2,k_1^2,k_2^2, m_i^2, m_i^2, m_j^2].
\end{align}
The tensor coefficients $C_i$ and $C_{ij}$ are defined according to the {\tt FeynCalc} convention~\cite{Mertig:1990an}.
\section{Counter terms for vertex renormalization}\label{app:cts}
\noindent In our model, the contributions to the counter terms in Eqs.~(\ref{eq:counterv}) and (\ref{eq:counterh}) are given by
\begin{align}
\delta M_W^2 & = \frac{1}{8\pi^2}\sum_{i=1,2,3} g_W^{ic} g_W^{ci} C_V(m_i,m_c), \quad \delta Z_W = \frac{1}{24\pi^2}\sum_{i=1,2,3} g_W^{ic} g_W^{ci} C_{\partial V}(m_i,m_c),\\\nonumber
\delta M_Z^2 & = \frac{1}{8\pi^2}\Big[(g_Z^{cc})^2 C_V(m_c,m_c) + \sum_{i=1,3} g_Z^{i2} g_Z^{2i} C_V(m_i,m_2) \Big], \\\nonumber
\delta Z_{Z} & = \frac{1}{24\pi^2}\Big[(g_Z^{cc})^2 C_{\partial V}(m_c,m_c) + \sum_{i=1,3} g_Z^{i2} g_Z^{2i} C_{\partial V}(m_i,m_2) \Big],\\\nonumber
\delta Z_e & = - \frac{\delta Z_{AA}}{2} - \frac{s_W}{c_W}\frac{\delta Z_{ZA}}{2},\quad \delta Z_{AA} = - \frac{e^2}{12\pi^2}B_0[0,m_c^2,m_c^2],\quad \delta Z_{ZA} = 0,\quad \frac{\delta s_W}{s_W} = -\frac{c_W^2}{2s_W^2}\Big(\frac{\delta M_W^2}{M_W^2} - \frac{\delta M_Z^2}{M_Z^2}\Big),\\\nonumber
\delta M_h^2 & = \frac{1}{16\pi^2}\sum_{ij=11,13,31,33} (g_h^{ij})^2 C_h(m_i,m_j),\quad \delta Z_h = \frac{1}{16\pi^2}\sum_{ij=11,13,31,33} (g_h^{ij})^2 C_{\partial h}(m_i,m_j),\\\nonumber
\delta t & = \frac{i}{8\pi^2}\sum_{i=1,3} g_h^{ii}\,m_i A_0(m_i^2).
\end{align}
The loop functions for the counter terms are defined by
\begin{align}
C_V(m_i,m_j) & = A_0[m_i^2] + A_0[m_j^2] + \big((m_j - m_i)^2 - M_V^2\big) B_0[M_V^2, m_i^2, m_j^2] - 4B_{00}[M_V^2, m_i^2, m_j^2],\\\nonumber
C_{\partial V}(m_i,m_j) & =  \frac{(m_j^2 - m_i^2)^2}{M_V^4}\Big[B_0[M_V^2, m_i^2, m_j^2] - B_0[0, m_i^2, m_j^2]\Big]
+ 2B_0[M_V^2, m_i^2, m_j^2]\\\nonumber
 & \qquad\, - \big((m_j^2 - m_i^2)^2/M_V^2 + m_j^2 - 6m_i m_j + m_i^2 - 2M_V^2 \big)\partial B_0[M_V^2, m_i^2, m_j^2] - \frac{2}{3},\\\nonumber
 C_h(m_i,m_j) & = A_0[m_i^2] + A_0[m_j^2] + \big((m_i + m_j)^2 - M_h^2\big)B_0[M_h^2, m_i^2, m_j^2],\\\nonumber
 C_{\partial h}(m_i,m_j) & = B_0[M_h^2, m_i^2, m_j^2] - \big((m_i + m_j)^2 - M_h^2\big)\partial B_0[M_h^2, m_i^2, m_j^2].
\end{align}
Here $\partial B_0[M^2,m_i^2,m_j^2]$ denotes the partial derivative of $B_0[p^2,m_i^2,m_j^2]$ with respect to $p^2$, evaluated at $p^2=M^2$.

\end{document}